\documentclass[a4paper,11pt]{article}
\pdfoutput=1

\usepackage{jheppub}
\usepackage[T1]{fontenc}
\usepackage{caption}
\usepackage{subcaption}
\usepackage{float}

\usepackage{booktabs} 

\RequirePackage[dvipsnames]{xcolor}

\newcommand{\jewel}{\textbf{\textsc{Jewel}}}

\title{\boldmath All is More: Energy Flow Networks for Jet Quenching}

\author[a,b]{J. A. Gonçalves}
\affiliation[a]{Laboratório de Instrumentação e Física Experimental de Partículas (LIP), Av. Professor Gama Pinto 2, 1649-003 Lisboa, Portugal}
\affiliation[b]{Instituto Superior Técnico, Universidade de Lisboa, Av. Rovisco Pais 1, 1609-001, Lisboa, Portugal}
\emailAdd{jgponcalves@lip.pt}

\abstract{
Jet quenching, the modification of jets by the quark-gluon plasma in heavy-ion collisions, provides a sensitive probe of the properties of the medium. A jet-by-jet discrimination study between proton-proton and lead-lead jets using energy flow networks and simple baselines, explicitly retaining medium response and underlying event contamination is presented. As references, linear discriminants and neural networks have been trained on high-level observables such as $N$-subjettiness and energy flow polynomials, including an extended energy flow polynomial set, in order to quantify the achievable performance without constituent-level learning. Energy flow networks are then trained on jet constituents and extended to observable-enhanced energy flow networks that concatenate standardized $N$-subjettiness and/or energy flow polynomials to the energy flow network latent space. In the realistic scenario, including both underlying event contamination and medium response, observable-enhanced energy flow networks set state-of-the-art performance with receiver operating characteristic area under the curve of $\simeq 0.83$, improving markedly over linear and non-linear baselines and previous work with different architectures. Finally, results from moment energy flow networks, an energy flow network variant that attains comparable area under the curve with a substantially more compact and interpretable latent space are shown. These results establish energy-flow-network-based approaches (especially when enhanced with physics-motivated observables) as practical and robust tools for jet-quenching studies.
}

\begin{document}
\maketitle
\flushbottom

\section{Introduction}
\label{sec:intro}

Jets produced in ultrarelativistic heavy-ion (HI) collisions traverse and interact with the quark-gluon plasma (QGP), undergoing medium-induced modifications collectively known as jet quenching~\cite{Cao:2024pxc}. These modifications encode space-time information about the plasma and its transport properties, motivating per-jet methods that can identify, jet-by-jet, which reconstructed jets are substantially modified and which remain effectively vacuum-like~\cite{Busza:2018rrf,Apolinario:2022vzg,Cao:2024pxc,Cunqueiro:2021wls,Vertesi:2024tdv}. As a practical proxy for per-jet modification, a binary classifier is trained to separate jets from proton-proton (pp) events --- taken as vacuum-like --- from jets in PbPb events, which contain an admixture of vacuum-like and quenched jets. This surrogate is not identical to a ``quenched vs.\ unquenched'' label, but it is a controlled and measurable step toward jet-by-jet quenching assessments. 

Recent machine learning (ML) work has trained supervised classifiers and regressors to expose quenching on a per-jet basis --- predicting fractional energy loss and even localizing dijet production points in the nuclear overlap region~\cite{Du:2020pmp, Du:2021pqa} --- while complementary studies build calculable, interpretable discriminants and rank which infrared and collinear (IRC) safe or unsafe features carry the most separating power~\cite{Apolinario:2021olp, Lai:2021ckt}. Quenched-jet taggers trained on JEWEL vs. PYTHIA quantify discrimination and its degradation under underlying event (UE) and subtraction, and show strong gains when medium response (MR) is present~\cite{Liu:2022hzd, Qureshi:2024ceh}. A machine-learning-driven scan of jet-substructure observables further identifies robust, portable choices for experimental use~\cite{CrispimRomao:2023ssj}. To prevent background-driven shortcuts, an ``apples-to-apples'' baseline --- pp jets embedded in a realistic heavy-ion UE and subtracted identically to PbPb --- has been advocated so any residual differences can be cleanly attributed to genuine quenching~\cite{ArrudaGoncalves:2025wtb}. For a broader perspective on machine learning in particle physics, see the living review~\cite{hepmllivingreview}.

Realistic interpretation demands that two effects be handled explicitly. MR, i.e.\ recoil partons induced by jet-medium interactions, populates soft, often large-angle radiation and can become an unphysical shortcut for discrimination if mishandled. In \jewel, recoils are therefore retained and the dedicated thermal-recoil subtraction procedure that has been validated for jet-shape systematics~\cite{KunnawalkamElayavalli:2017hxo} is applied. In parallel, the heavy-ion UE degrades discrimination and can mask genuine quenching features; to avoid background-driven artifacts the ``apples-to-apples'' prescription~\cite{ArrudaGoncalves:2025wtb} has been followed, embedding the pp reference in a realistic HI UE and treating subtraction analogously to the PbPb case so that any residual differences are attributable to quenching rather than UE contamination.

Our approach combines physics-motivated observables with constituent-level learning. We examine several observables, namely $N$-subjettiness bases with \(N\) up to \(21\), according to Refs.~\cite{Thaler:2010tr,Thaler:2011gf} given by:
\begin{equation}
\left\{ \tau^{1/2}_{1},\, \tau^{1}_{1},\, \tau^{2}_{1},\, 
\tau^{1/2}_{2},\, \tau^{1}_{2},\, \tau^{2}_{2},\, \ldots,\,
\tau^{1/2}_{N-2},\, \tau^{1}_{N-2},\, \tau^{2}_{N-2},\,
\tau^{1}_{N-1},\, \tau^{2}_{N-1} \right\}
\end{equation}
and, to span multiparticle energy-angle correlations with a complete linear IRC-safe basis, energy flow polynomials (EFPs)~\cite{Komiske:2017aww}, given by:
\begin{equation}
\label{eq:efp}
\text{EFP}_G = \sum_{i_1, \dots, i_N} z_{i_1} \dots z_{i_N} \prod_{(k,\ell) \in G} \theta_{i_k i_\ell},
\end{equation}
where \(G\) is a given graph structure, \(z_i\) are transverse-momentum fractions defined as \(z_i = p_{T,i}/p_T^{\text{jet}}\) with \(\sum_i z_i = 1\), and \(\theta_{ij}=\sqrt{(\Delta\eta_i-\Delta\eta_j)^2+(\Delta\phi_i-\Delta\phi_j)^2}\) are distances in pseudorapidity-azimuth space. An extended variant was also considered with a set of exponents for \(z\) and \(\theta\). On the constituent-learning side, energy flow networks (EFNs) were adopted, which specialize deep sets to collider data and realize IRC safety at the representation level through transverse-momentum-fraction-weighted, permutation-invariant pooling over latent space variables~\cite{Komiske:2018cqr,Zaheer:2017wmg}. These networks can be mathematically described as:
\begin{equation}
\label{eq:efn}
\text{EFN}: F\left(\sum_{i} z_i \, \Phi(\hat{p}_i)\right),
\end{equation}
where \( \Phi: \mathbb{R}^2 \to \mathbb{R}^L \) is a learned per-particle mapping into a latent space of dimension \( L \), \( F: \mathbb{R}^L \to \mathbb{R}^k \) is a second neural network acting on the aggregated representation and \(\hat{p}_i=(\Delta\eta_i,\Delta\phi_i)\) are constituent cylindrical detector coordinates relative to the jet axis.

Two further EFN variants have been considered. Observable-enhanced EFNs (oEFNs), introduced in this paper, concatenate standardized high-level observables (e.g.\ $N$-subjettiness and/or EFPs) to the EFN latent space summary, injecting domain knowledge into an IRC-aware architecture. Moment EFNs (MEFNs) replace the sum in Eq.~\ref{eq:efn} by moment pooling defined as:
\begin{equation}
\label{eq:mefn}
\text{MEFN}: F_k \left( \langle \Phi^a \rangle_P, \langle \Phi^{a_1} \Phi^{a_2} \rangle_P, \dots, \langle \Phi^{a_1} \dots \Phi^{a_k} \rangle_P \right),
\end{equation}
where \(k\) is the highest order moment considered, and \(a\) are indices running through all latent space dimensions. This results in a significant increase in the input dimension of the \(F\) network, \(L_{\mathrm{eff}}\) relative to the chosen latent space dimension. In fact:
\begin{equation}
    L_{\mathrm{eff}} = \binom{L + k}{k}.
\end{equation}
\noindent This generalizes the expectation value of the EFN from a simple average to a vector of the moments of the latent space. These models, exposing these moments of the latent features, yield compact and more interpretable representations while maintaining competitive performance~\cite{Gambhir:2024dtf}. Where appropriate, weak supervision results via classification without labels (CWoLa) are reported to reflect the experimental reality that definitive ``quenching labels'' are currently unavailable~\cite{Metodiev:2017vrx}.

Hadron-level jets from \jewel~2.3.0 with MR retained were considered and an experimentally motivated UE, as well as correct UE handling under the apples-to-apples prescription~\cite{ArrudaGoncalves:2025wtb} was employed. Linear and non-linear baselines have been established on $N$-subjettiness, EFPs, and extended EFP sets. EFNs trained on constituents were shown to significantly outperform observable-only baselines in the realistic (UE+MR) regime. Furthermore, oEFNs are shown to deliver further, robust gains. MEFNs that match EFN-level AUC with markedly smaller, interpretable latent spaces, were trained, with their observable-enhanced counterparts (oMEFNs) reaching oEFN-level performance. Throughout, $K$-fold averaged ROC AUC with cross-validation was used as the primary performance metric and the standard deviation across folds as the primary uncertainty metric. In the UE+MR setting, our best oEFN/oMEFN configurations set state-of-the-art per-jet discrimination consistent with, and improving upon, the apples-to-apples benchmark~\cite{ArrudaGoncalves:2025wtb}. Sec.~\ref{sec:simul} summarizes the simulated samples, recoil subtraction, and UE handling. Sec.~\ref{sec:analy} presents observable diagnostics, linear/non-linear baselines, EFN/oEFN results, and MEFN studies. Conclusions follow in Sec.~\ref{sec:concl}.

\section{Simulation setup}
\label{sec:simul}

All results use hadron-level jets from \jewel~2.3.0~\cite{Zapp:2013vla} at $\sqrt{s_{NN}}=5.02$~TeV with MR and an experimentally motivated UE model, following the same samples and processing as Ref.~\cite{ArrudaGoncalves:2025wtb}, where the UE is subtracted using iterative constituent subtraction (ICS)~\cite{Berta:2014eza}. Baselines are constructed in two regimes: (i) ``no UE'' (optimistic ceiling) and (ii) ``with UE contamination'' (realistic), and focus on the latter for the main results. Main results come from classifiers trained with identical $K$-fold splits.

Two samples were generated with identical hard-process settings: pp and PbPb in the \([0,10]\%\) centrality bin. In both cases, jets are reconstructed at hadron level after hadronization, with \jewel\ recoils enabled and kept for subsequent subtraction (thermal-recoil subtraction validated for \jewel)~\cite{Milhano:2022kzx}. Jets are clustered using the anti-$k_t$ algorithm with jet parameter $R=0.4$ as implemented in \textsc{FastJet}~\cite{Cacciari:2011ma,Cacciari:2008gp}. Both charged and neutral particles are included, and no grooming is applied. 

For observable baselines, $N$-subjettiness and EFPs are computed on the subtracted constituents. The extended EFPs use $(\kappa,\beta)\in\{0.5,1\}\times\{0.5,1,2\}$ on the same constituent set~\cite{Komiske:2017aww}. For EFN-based models, inputs are the constituent transverse-momentum fractions $z_i=p_{T,i}/p_T^{\text{jet}}$ and angular coordinates, namely translated pseudorapidity-azimuth, $(\Delta \eta_i,\Delta\phi_i)$ relative to the jet axis; jets are centered to $(0,0)$ in $(\eta,\phi)$ with $\phi\in(-\pi,\pi]$, and no additional per-jet rescaling beyond $z_i$ normalization is applied.

Evaluation uses $K$-fold cross-validation with fixed, splits shared across all models and input sets to enable one-to-one comparisons. Unless noted otherwise, headline results correspond to the realistic UE+MR configuration; the no UE configuration is used only as a guiding baseline. Where indicated, weak supervision via CWoLa (see Ref.~\cite{Metodiev:2017vrx}) is performed by considering PbPb jets as a mixture of a vacuum-like distribution and a quenched-jet distribution, while pp jets with UE contamination are taken to be a pure sample of vacuum-like jets, equivalent to the vacuum-like component of the PbPb case.

All observable inputs are standardized using training-fold statistics. The non-linear model is a small one-layer neural network (NN); oEFN concatenates the standardized observable vector to the EFN latent before $F$. MEFN replaces sum pooling by moments up to order $k$; oMEFN concatenates observables analogously. All networks are optimized using Adam~\cite{kingma2014adam}, early stopping on validation loss, and fixed seeds. Evaluation uses identical $K$-fold splits shared across all models. The neural architectures and training hyperparameters used throughout are summarized in Tab.~\ref{tab:arches}. Quoted uncertainties are the standard deviation across all validation folds.

\begin{table}[!htbp]
\centering
\small
\begin{tabular}{lccccc}
\toprule
Model & Layers & Activation & Patience & Dropout & $L_2$ \\
\midrule
NN & (2048) & ReLU & 30 & 0.20 & 0.005 \\
EFN & $\Phi$: (100, 100, 126), $F$: (100, 100, 100) & ReLU & 30 & 0.075 & --- \\
oEFN & $\Phi$: (100, 100, 126), $F$: (100, 100, 100) & ReLU & 30 & 0.20 & --- \\
MEFN & $\Phi$: (100, 100, $L$), $F$: (100, 100, 100) & ReLU & 30 & 0.20 & --- \\
oMEFN & $\Phi$: (100, 100, $L$), $F$: (100, 100, 100) & ReLU & 30 & 0.20 & --- \\
\bottomrule
\end{tabular}
\caption{Network architectures and training hyperparameters. ``Patience'' is the early-stopping patience (epochs without validation loss improvement). For MEFN, $L$ denotes the latent size scanned in Sec.~\ref{sssec:mefn}.}
\label{tab:arches}
\end{table}

\section{Results}
\label{sec:analy}

\subsection{Observables: distributions and correlations}
\label{ssec:obs_diagnostics}

Before turning to physics-informed architectures, baseline performance is established using high-level observable sets ($N$-subjettiness, EFPs, and an extended EFP basis). These baselines serve as reference points and sanity checks for discrimination power under controlled complexity and are evaluated both without and with UE contamination for completeness.

\begin{figure}[!htbp]
    \centering
    \begin{subfigure}[b]{\textwidth}
        \includegraphics[width=\textwidth]{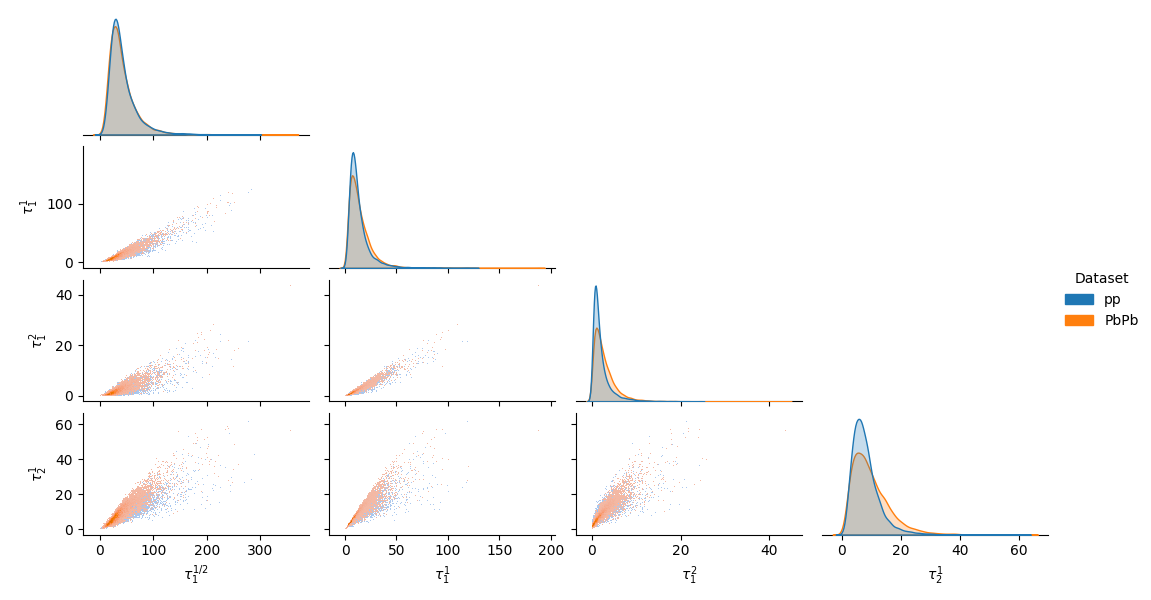}
    \end{subfigure}
    \begin{subfigure}[b]{\textwidth}
        \includegraphics[width=\textwidth]{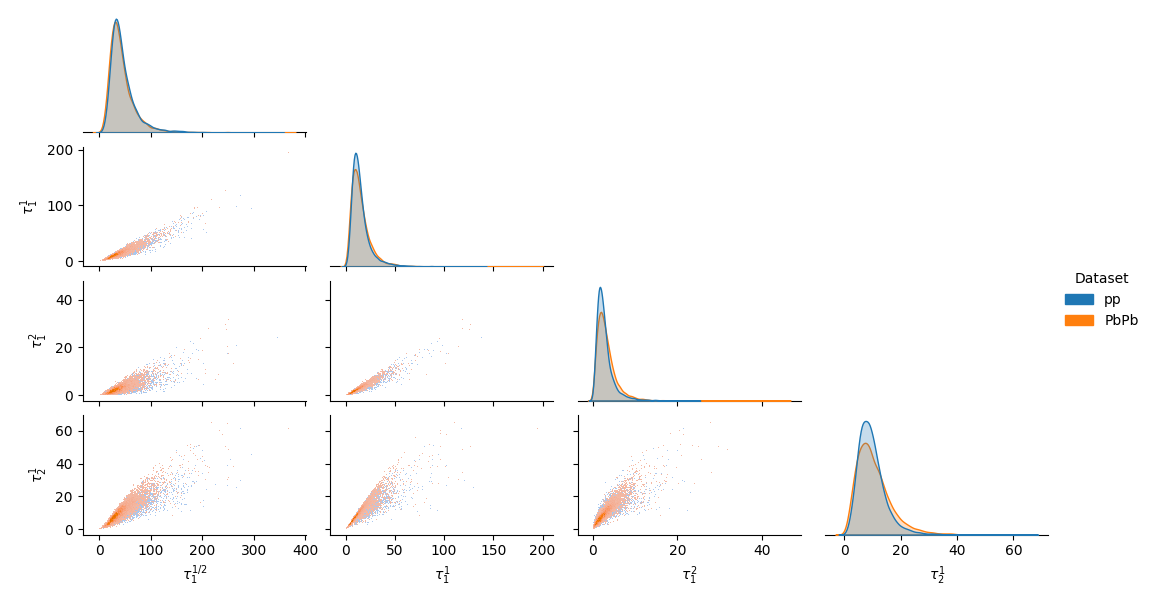}
    \end{subfigure}
    \caption{Distributions of 4 $N$-subjettiness variables and their correlations (pairwise scatter with marginal histograms); (top) without UE contamination; (bottom) with UE contamination.}
    \label{fig:baseline_overview_nsubs}
\end{figure}

Figure~\ref{fig:baseline_overview_nsubs} shows the distributions and pairwise correlations of four $N$-subjettiness observables, $\tau^{1/2}_{1}$, $\tau^{1}_{1}$, $\tau^{2}_{1}$, and $\tau^{1}_{2}$, for jets reconstructed in pp and PbPb samples. In the absence of UE contamination, the marginal distributions display slight differences between the two collision systems for most variables, with the notable exception of $\tau^{1/2}_{1}$, whose distributions remain nearly identical. Despite this similarity, correlations of $\tau^{1/2}_{1}$ with the remaining observables still reveal discernible separation between pp and PbPb jets, a trend that is consistently observed across all correlation panels. Once UE contamination is included, the distributions become more similar across the two systems, with very slightly broader shapes for pp and slightly narrower for PbPb with an accompanying reduced difference in the maximum of the distributions, indicating that background activity partially smears out the quenching-driven differences. This seems to indicate that discriminatory power in $N$-subjettiness arises less from individual observables and slightly more from the correlated patterns among them, and that these correlations are less vulnerable to soft contamination, when compared to the absolute distributions.

\begin{figure}[!htbp]
    \centering
    \begin{subfigure}[b]{\textwidth}
        \includegraphics[width=\textwidth]{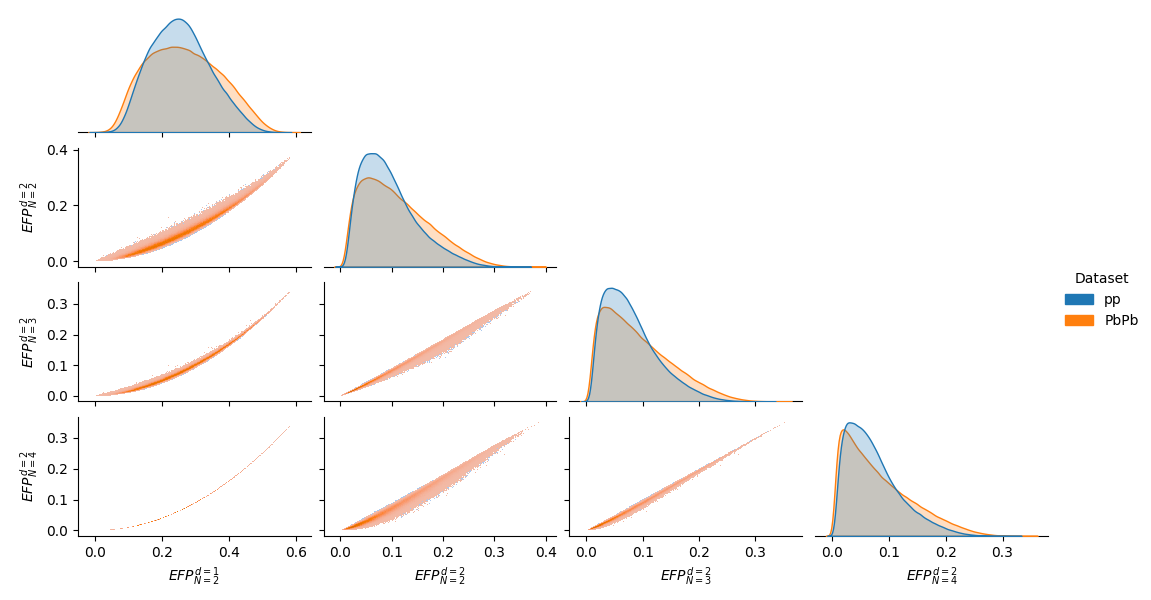}
    \end{subfigure}
    \begin{subfigure}[b]{\textwidth}
        \includegraphics[width=\textwidth]{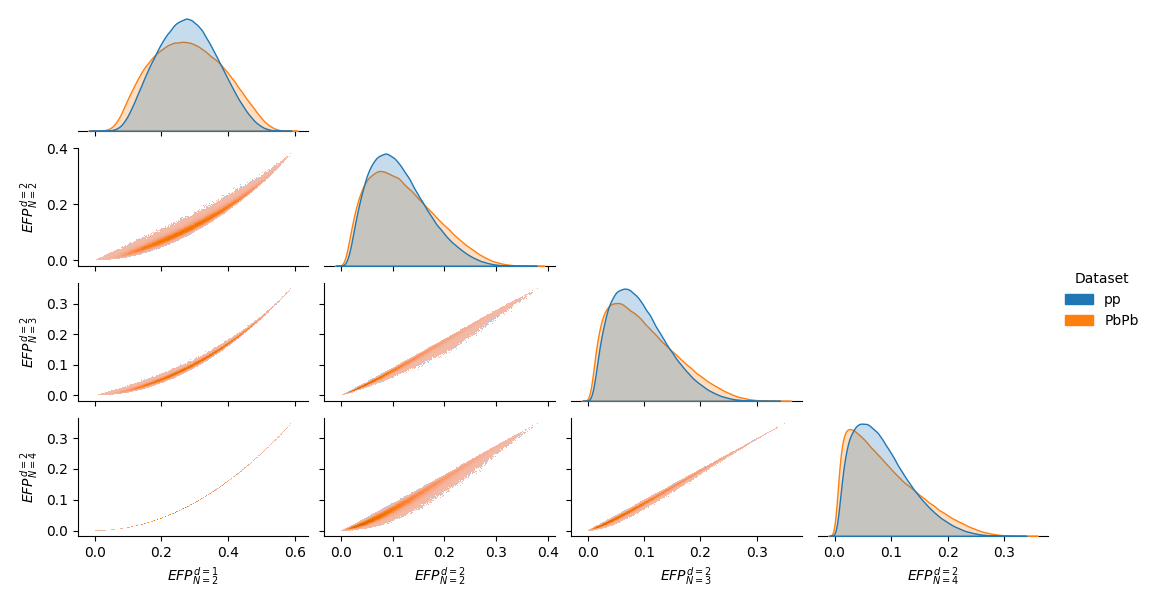}
    \end{subfigure}
    \caption{Distributions of 4 EFPs and their correlations (pairwise scatter with marginal histograms); (top) without UE contamination; (bottom) with UE contamination.}
    \label{fig:baseline_overview_efps}
\end{figure}

Figure~\ref{fig:baseline_overview_efps} presents the distributions and pairwise correlations of four selected EFPs: $EFP^{d=1}_{N=2}$, $EFP^{d=2}_{N=2}$, $EFP^{d=2}_{N=3}$, and $EFP^{d=2}_{N=4}$, as defined in Eq.~\ref{eq:efp}. In contrast to the $N$-subjettiness case, the marginal distributions of all four observables are shown to be more distinct between the pp and PbPb samples, both with and without UE contamination. The correlations, however, tend to overlap more strongly, suggesting that while each EFP{\parfillskip=0pt\par} 
\begin{figure}[H]
\centering
    \begin{subfigure}[b]{0.45\textwidth}
        \centering
        \includegraphics[width=\textwidth]{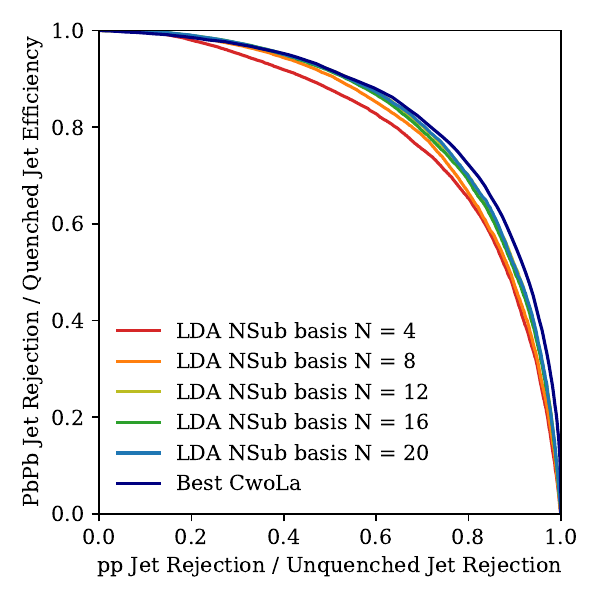}
        \caption{no UE, $N$-subjettiness}
        \label{fig:roc_nsub_noue}
    \end{subfigure}
    \hfill
    \begin{subfigure}[b]{0.45\textwidth}
        \centering
        \includegraphics[width=\textwidth]{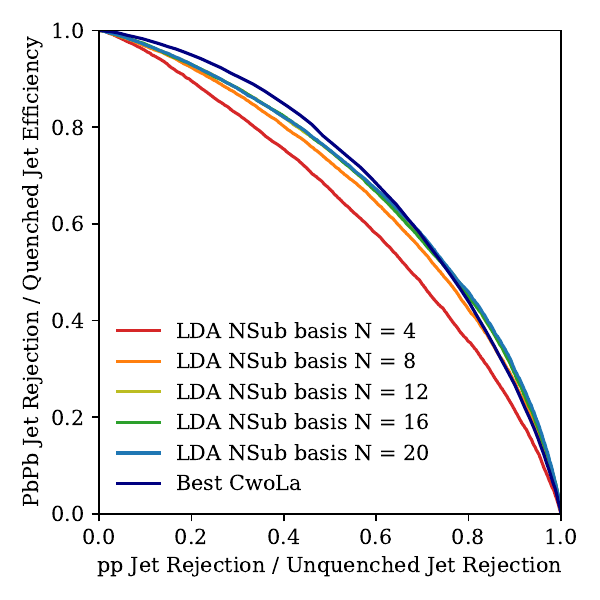}
        \caption{with UE, $N$-subjettiness}
        \label{fig:roc_nsub_ue}
    \end{subfigure}

    \begin{subfigure}[b]{0.45\textwidth}
        \centering
        \includegraphics[width=\textwidth]{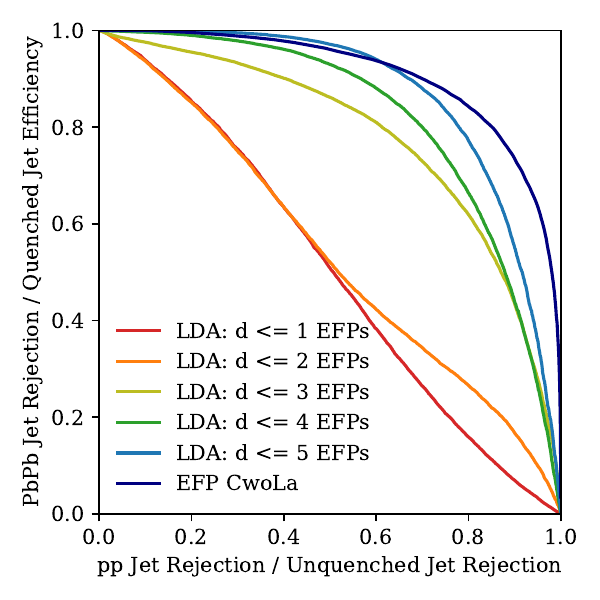}
        \caption{no UE, EFPs}
        \label{fig:roc_efp_noue}
    \end{subfigure}
    \hfill
    \begin{subfigure}[b]{0.45\textwidth}
        \centering
        \includegraphics[width=\textwidth]{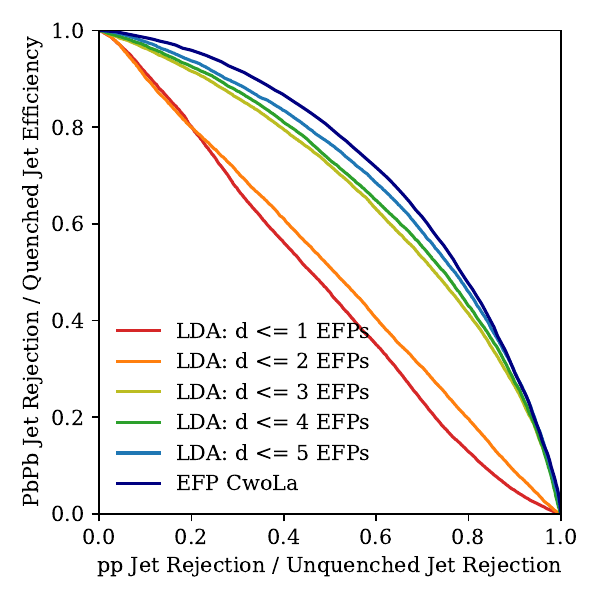}
        \caption{with UE, EFPs}
        \label{fig:roc_efp_ue}
    \end{subfigure}
    \begin{subfigure}[b]{0.45\textwidth}
        \centering
        \includegraphics[width=\textwidth]{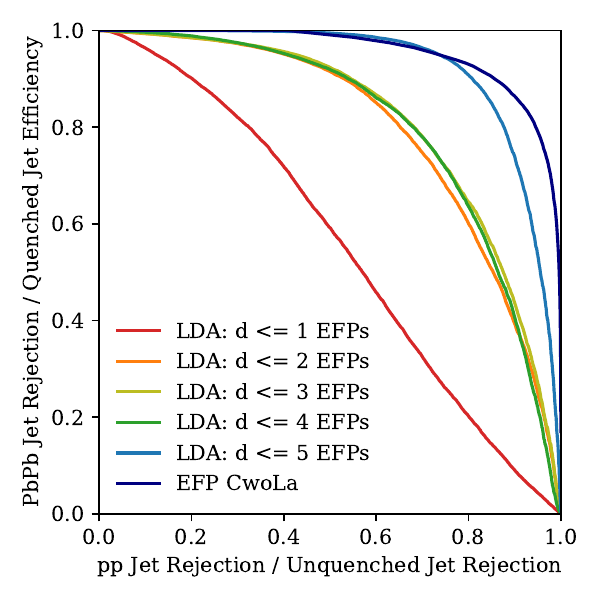}
        \caption{no UE, extended EFPs}
        \label{fig:roc_efpext_noue}
    \end{subfigure}
    \hfill
    \begin{subfigure}[b]{0.45\textwidth}
        \centering
        \includegraphics[width=\textwidth]{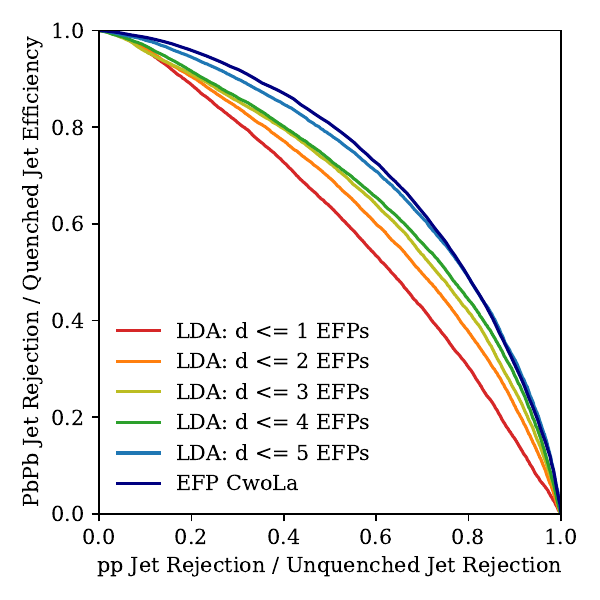}
        \caption{with UE, extended EFPs}
        \label{fig:roc_efpext_ue}
    \end{subfigure}

    \caption{ROC curves for linear models across input sets and UE configurations.}
    \label{fig:linear_roc_six}
\end{figure}
\begin{figure}[!h]
    \centering
    \begin{subfigure}[b]{0.49\textwidth}
        \centering
        \includegraphics[width=\textwidth]{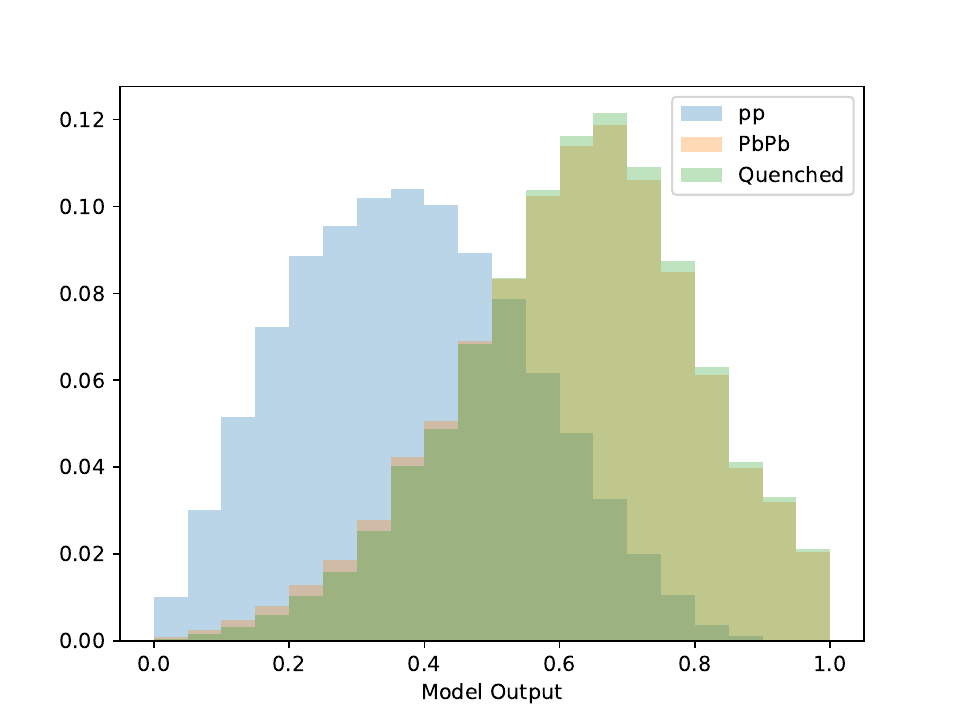}
        \caption{no UE, $N$-subjettiness}
        \label{fig:out_nsub_noue}
    \end{subfigure}
    \hfill
    \begin{subfigure}[b]{0.49\textwidth}
        \centering
        \includegraphics[width=\textwidth]{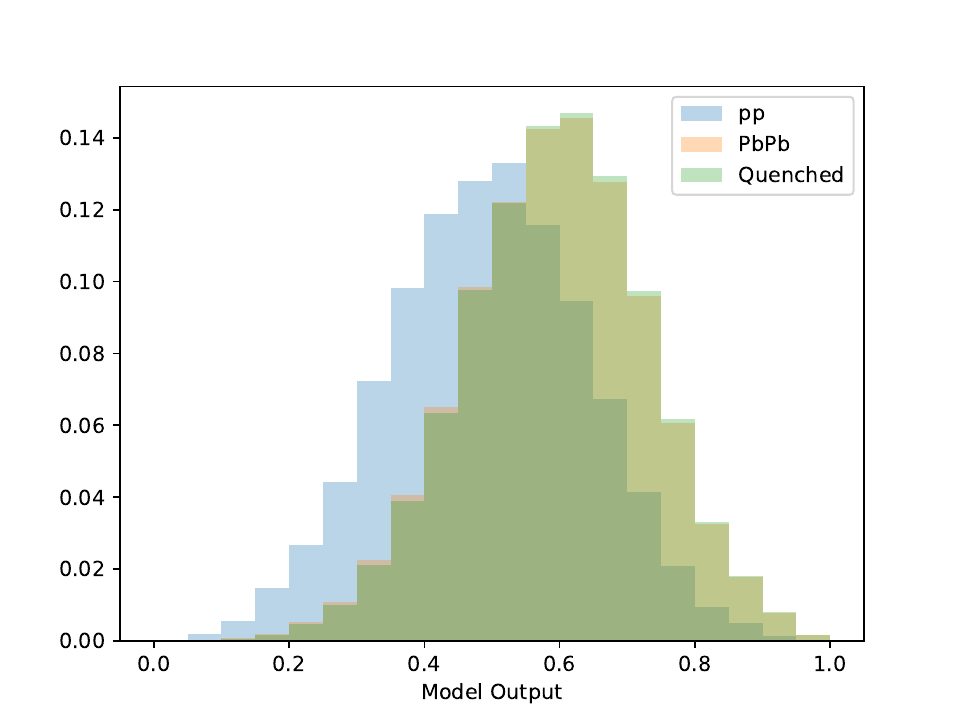}
        \caption{with UE, $N$-subjettiness}
        \label{fig:out_nsub_ue}
    \end{subfigure}
    \begin{subfigure}[b]{0.49\textwidth}
        \centering
        \includegraphics[width=\textwidth]{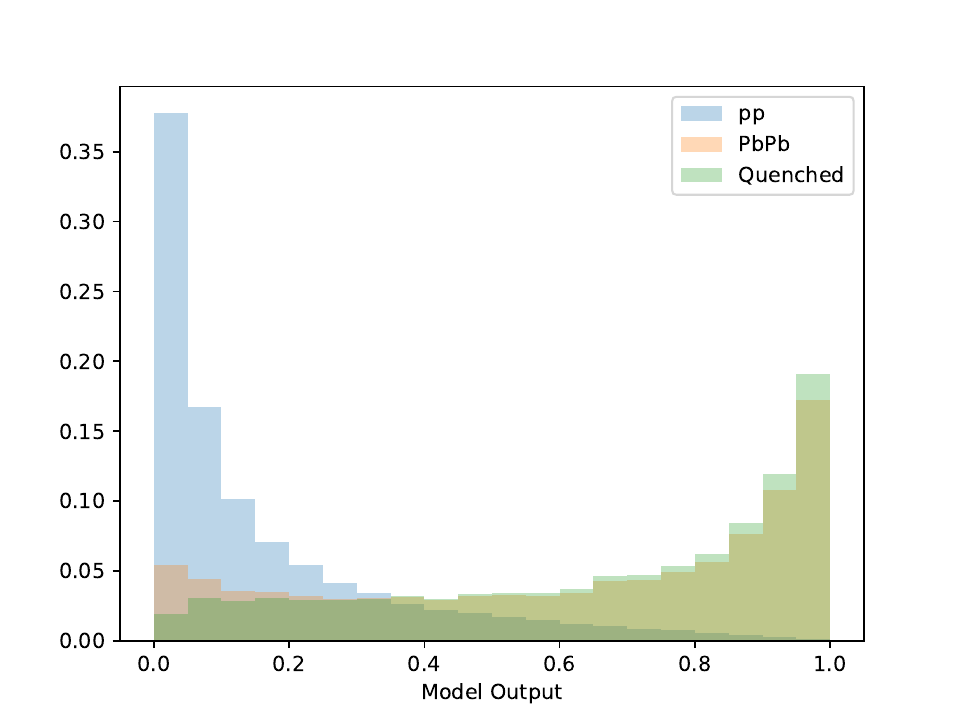}
        \caption{no UE, EFPs}
        \label{fig:out_efp_noue}
    \end{subfigure}
    \hfill
    \begin{subfigure}[b]{0.49\textwidth}
        \centering
        \includegraphics[width=\textwidth]{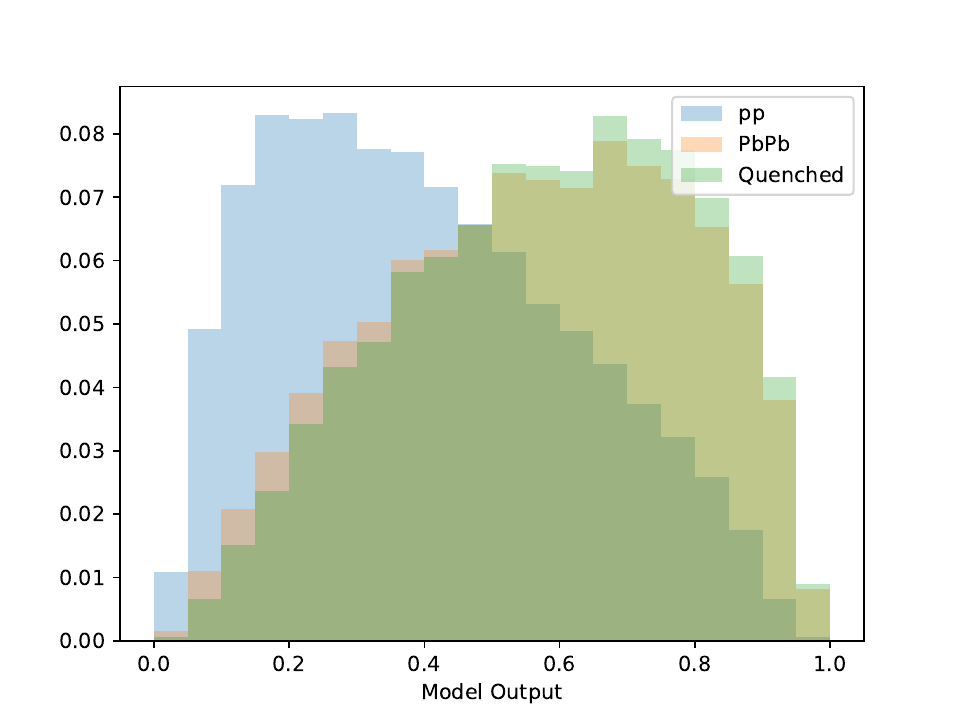}
        \caption{with UE, EFPs}
        \label{fig:out_efp_ue}
    \end{subfigure}
    \begin{subfigure}[b]{0.49\textwidth}
        \centering
        \includegraphics[width=\textwidth]{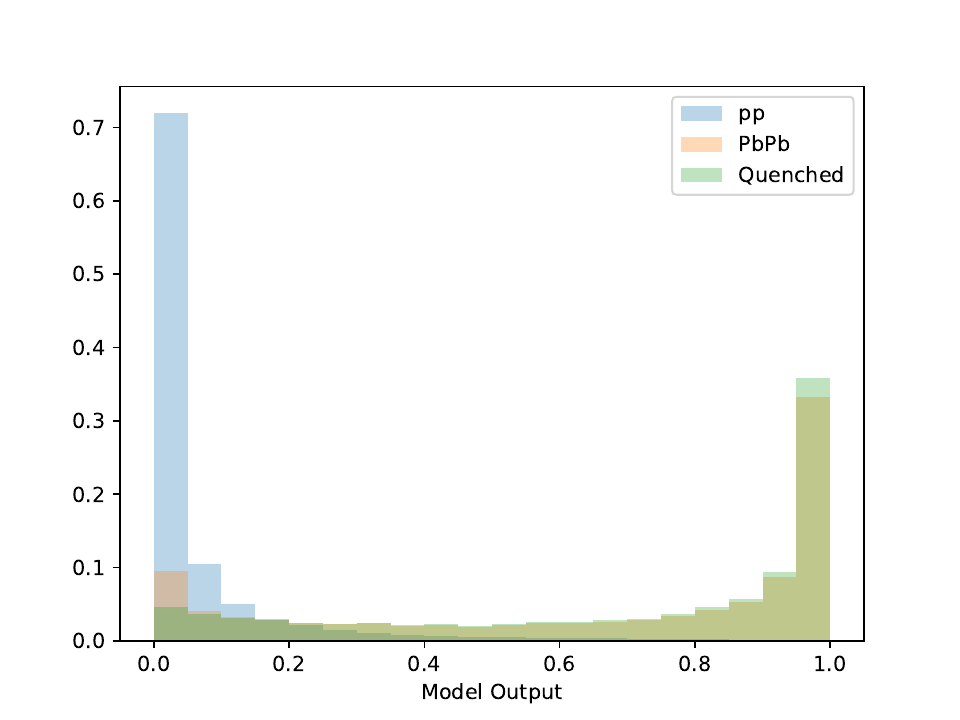}
        \caption{no UE, extended EFPs}
        \label{fig:out_efpext_noue}
    \end{subfigure}
    \hfill
    \begin{subfigure}[b]{0.49\textwidth}
        \centering
        \includegraphics[width=\textwidth]{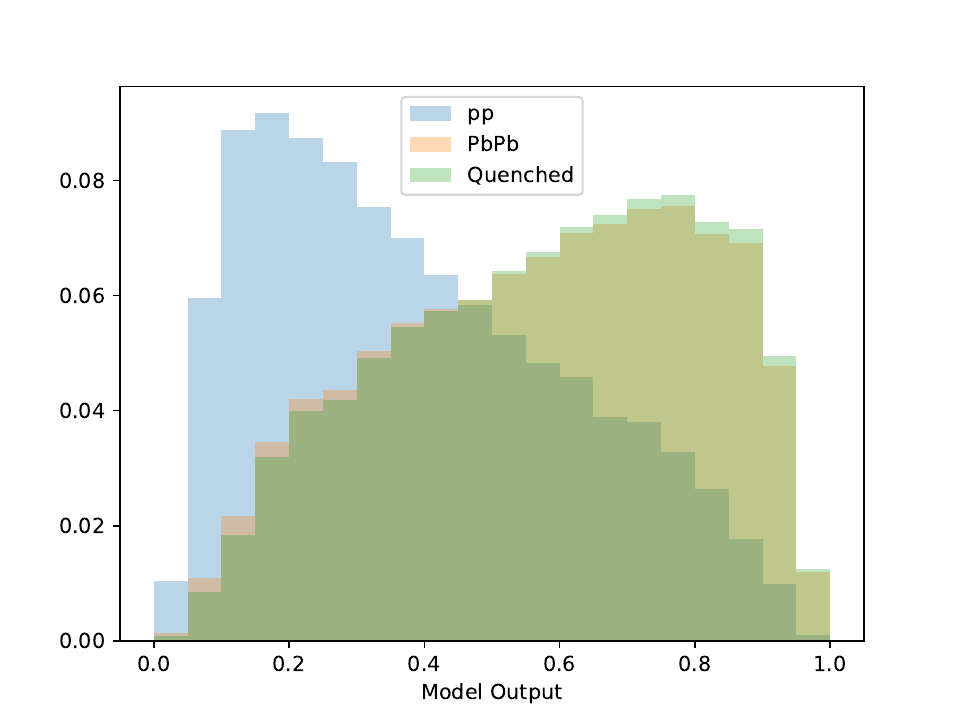}
        \caption{with UE, extended EFPs}
        \label{fig:out_efpext_ue}
    \end{subfigure}
    \caption{Classifier output score distributions for linear models across input sets and UE configurations.}
    \label{fig:linear_output_six}
\end{figure}
\noindent encodes quenching sensitivity individually, their joint structures are less distinctive across collision systems, at least at two-dimensional correlations. An interesting feature is the apparent non-linear one-to-one correspondence between $EFP^{d=1}_{N=2}$ and $EFP^{d=2}_{N=4}$, visible as a narrow curved band in their pairwise correlation panel. This relation reflects the polynomial hierarchy inherent in the EFP construction: for the specific graph choices used here, $EFP^{d=2}_{N=4}$ and $EFP^{d=1}_{N=2}$ are strongly correlated, with the former scaling approximately like the square of the latter.

\subsection{Linear baselines}
\label{ssec:linear}

The performance of linear classifiers trained on the different observable bases is summarized in Fig.~\ref{fig:linear_roc_six}. Linear discriminant analysis (LDA) was considered. For the $N$-subjettiness basis, the ROC curves in Figs.~\ref{fig:roc_nsub_noue} and~\ref{fig:roc_nsub_ue} show that increasing the number of observables yields only modest gains. In the absence of UE contamination, the presence of MR leads to a reasonably good separation (Fig.~\ref{fig:roc_nsub_noue}); once UE is included (Fig.~\ref{fig:roc_nsub_ue}), the curve moves toward the diagonal and the gap between smaller and larger $N$-subjettiness bases compresses. This pattern is consistent with the class-score behavior (Figs.~\ref{fig:out_nsub_noue} and~\ref{fig:out_nsub_ue}), where UE contamination reduces the mean separation between pp and PbPb scores.

For the EFP basis, Figs.~\ref{fig:roc_efp_noue} and~\ref{fig:roc_efp_ue} show decent discrimination in the no UE configuration --- slightly better than $N$-subjettiness at comparable operating points --- and a degradation under UE that nevertheless remains above the $N$-subjettiness curves. In contrast to $N$-subjettiness, including an increasing number of EFPs produces a clear and significant improvement of the ROC, reflecting the added value of higher-order angular correlations even for a linear decision boundary. The corresponding score distributions (Figs.~\ref{fig:out_efp_noue} and~\ref{fig:out_efp_ue}) retain a more pronounced valley between classes than in the $N$-subjettiness case, aligning with the stronger background rejection at fixed signal efficiency.

The extended EFPs, shown in Figs.~\ref{fig:roc_efpext_noue} and~\ref{fig:roc_efpext_ue}, generalize the exponents on momentum fractions and angular distances, $(\kappa,\beta)\in\{0.5,1\}\times\{0.5,1,2\}$, effectively scaling the basis size by a factor of six relative to the $(\kappa,\beta)=(1,1)$ set. This extension yields the best linear performance across both UE settings: a slight lift over standard EFPs without UE and a persistent advantage with UE. Notably, smaller subsets of the extended EFPs already match or exceed non-extended sets most likely due to their high-dimensionality, indicating that exponent tuning captures discriminating structure that simple basis growth in the $(1,1)$ family cannot --- again reflected in tighter and less-overlapping score densities (Figs.~\ref{fig:out_efpext_noue} and~\ref{fig:out_efpext_ue}).

\subsection{Non-linear baselines}
\label{ssec:nonlinear}

Linear decision boundaries are replaced with a feed-forward NN, trained on the same observable sets ($N$-subjettiness, EFPs, extended EFPs). This captures non-linear correlations among observables without leveraging low-level particle information. Models are trained and evaluated with consistent preprocessing and validation, both without and with UE contamination.

\begin{figure}[H]
    \centering    
    \begin{subfigure}[b]{0.45\textwidth}
        \centering
        \includegraphics[width=\textwidth]{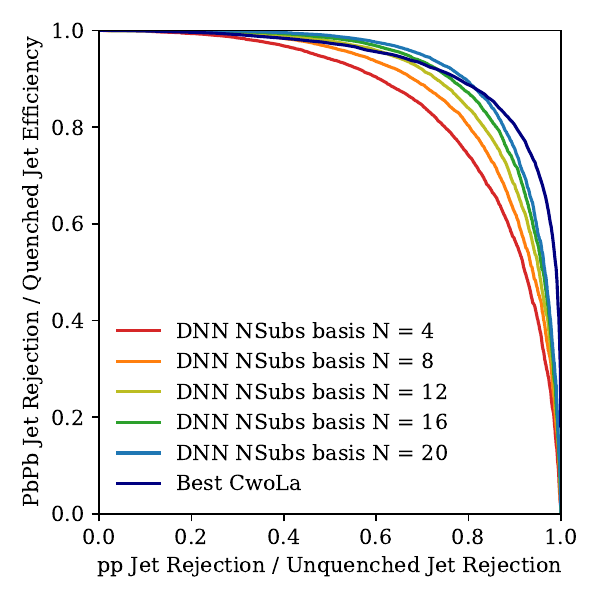}
        \caption{no UE, $N$-subjettiness}
        \label{fig:dnn_roc_nsub_noue}
    \end{subfigure}
    \hfill
    \begin{subfigure}[b]{0.45\textwidth}
        \centering
        \includegraphics[width=\textwidth]{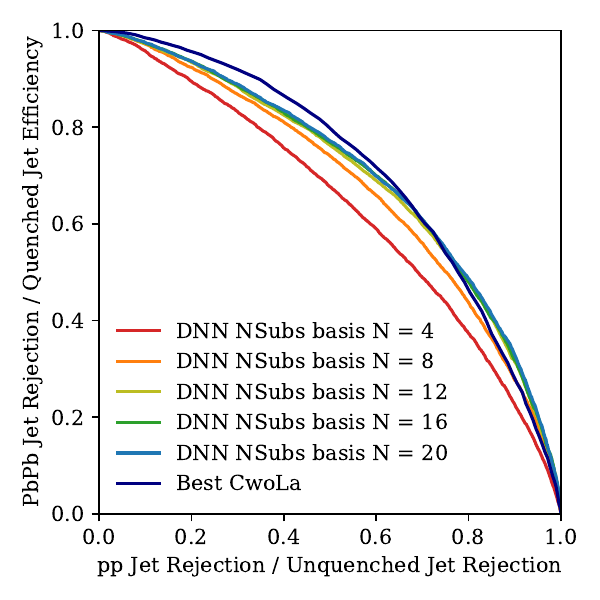}
        \caption{with UE, $N$-subjettiness}
        \label{fig:dnn_roc_nsub_ue}
    \end{subfigure}

    \begin{subfigure}[b]{0.45\textwidth}
        \centering
        \includegraphics[width=\textwidth]{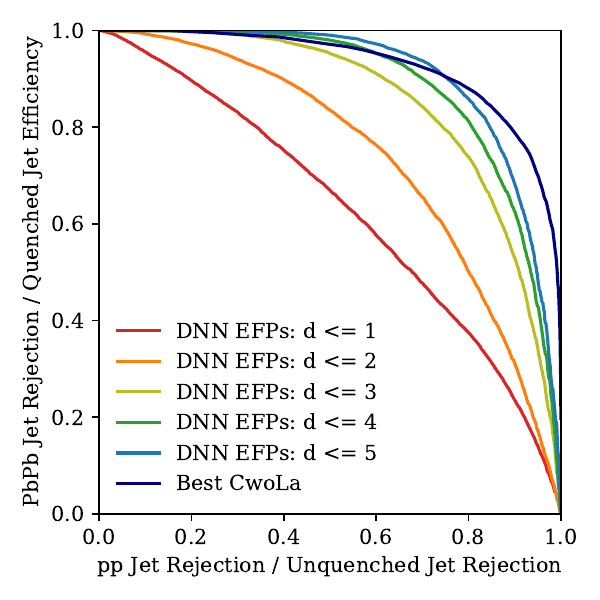}
        \caption{no UE, EFPs}
        \label{fig:dnn_roc_efp_noue}
    \end{subfigure}
    \hfill
    \begin{subfigure}[b]{0.45\textwidth}
        \centering
        \includegraphics[width=\textwidth]{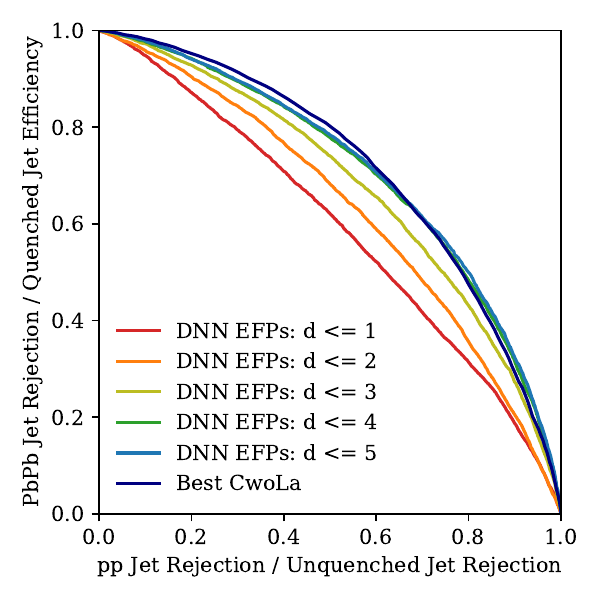}
        \caption{with UE, EFPs}
        \label{fig:dnn_roc_efp_ue}
    \end{subfigure}

    \begin{subfigure}[b]{0.45\textwidth}
        \centering
        \includegraphics[width=\textwidth]{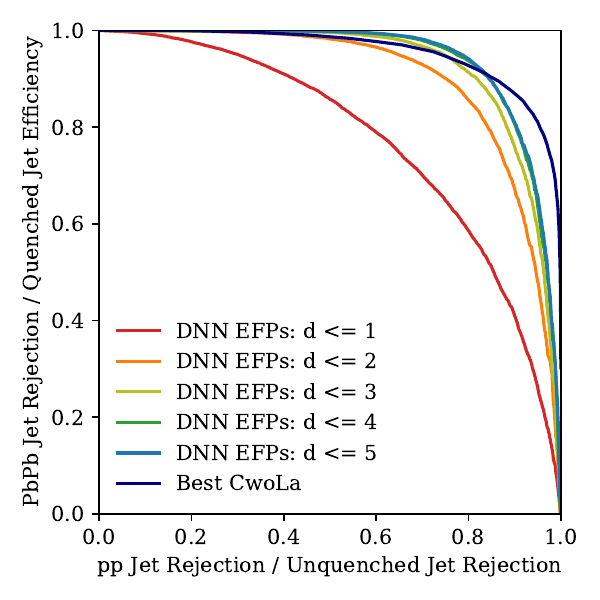}
        \caption{no UE, extended EFPs}
        \label{fig:dnn_roc_efpext_noue}
    \end{subfigure}
    \hfill
    \begin{subfigure}[b]{0.45\textwidth}
        \centering
        \includegraphics[width=\textwidth]{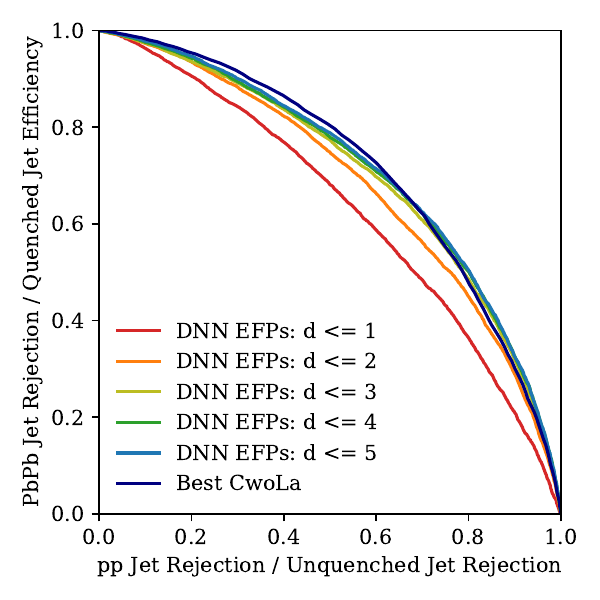}
        \caption{with UE, extended EFPs}
        \label{fig:dnn_roc_efpext_ue}
    \end{subfigure}

    \caption{ROC curves for NN classifiers across input sets and UE configurations.}
    \label{fig:dnn_roc_six}
\end{figure}

The ROC curves in Fig.~\ref{fig:dnn_roc_six} show the same ordering as in the linear case: the extended EFP basis provides the strongest discrimination, the standard EFP basis is next, and $N$-subjettiness performs the weakest of the three, although with comparable results. For $N$-subjettiness, Figs.~\ref{fig:dnn_roc_nsub_noue} and~\ref{fig:dnn_roc_nsub_ue}: in the no UE configuration the NN achieves a rather good separation, which is degraded but still above the state-of-the-art boosted decision tree{\parfillskip=0pt\par} 

\begin{figure}[!h]
    \centering
    \begin{subfigure}[b]{0.49\textwidth}
        \centering
        \includegraphics[width=\textwidth]{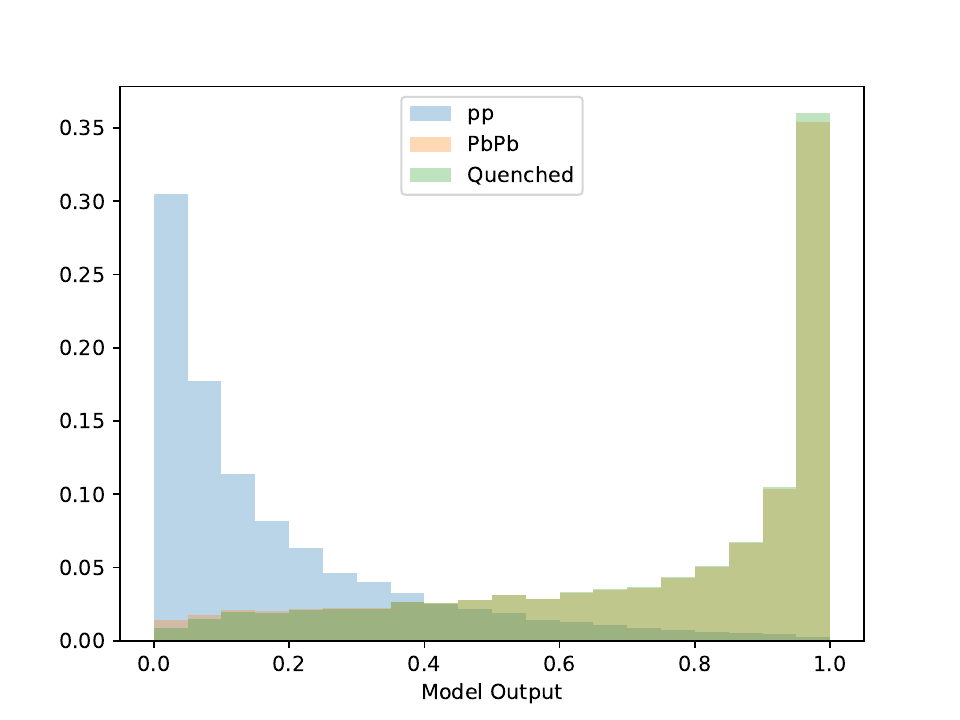}
        \caption{no UE, $N$-subjettiness}
        \label{fig:dnn_out_nsub_noue}
    \end{subfigure}
    \hfill
    \begin{subfigure}[b]{0.49\textwidth}
        \centering
        \includegraphics[width=\textwidth]{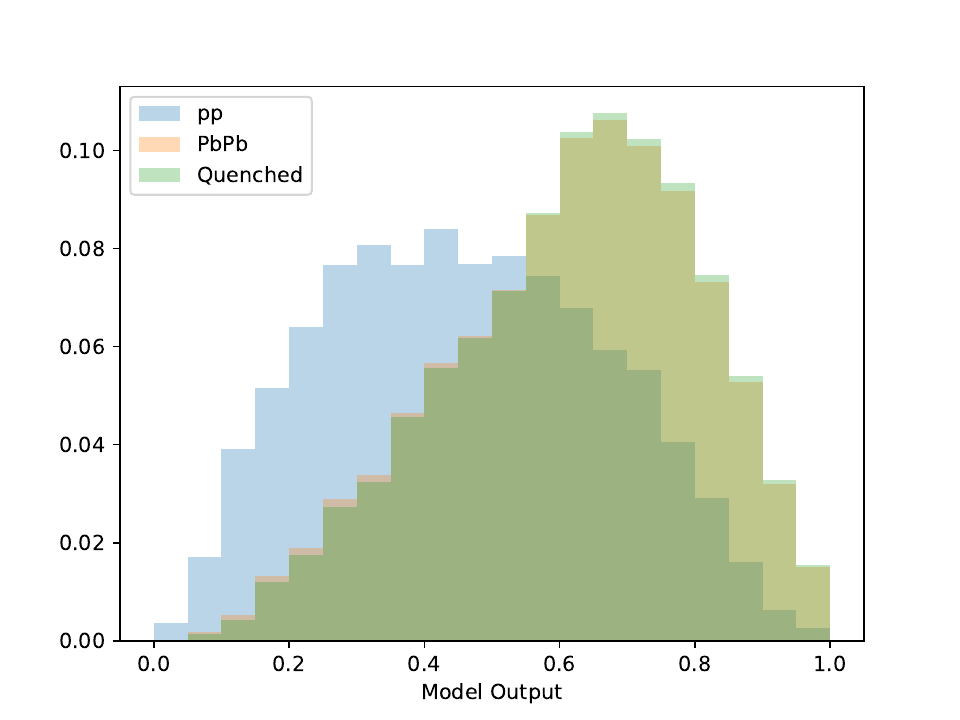}
        \caption{with UE, $N$-subjettiness}
        \label{fig:dnn_out_nsub_ue}
    \end{subfigure}

    \begin{subfigure}[b]{0.49\textwidth}
        \centering
        \includegraphics[width=\textwidth]{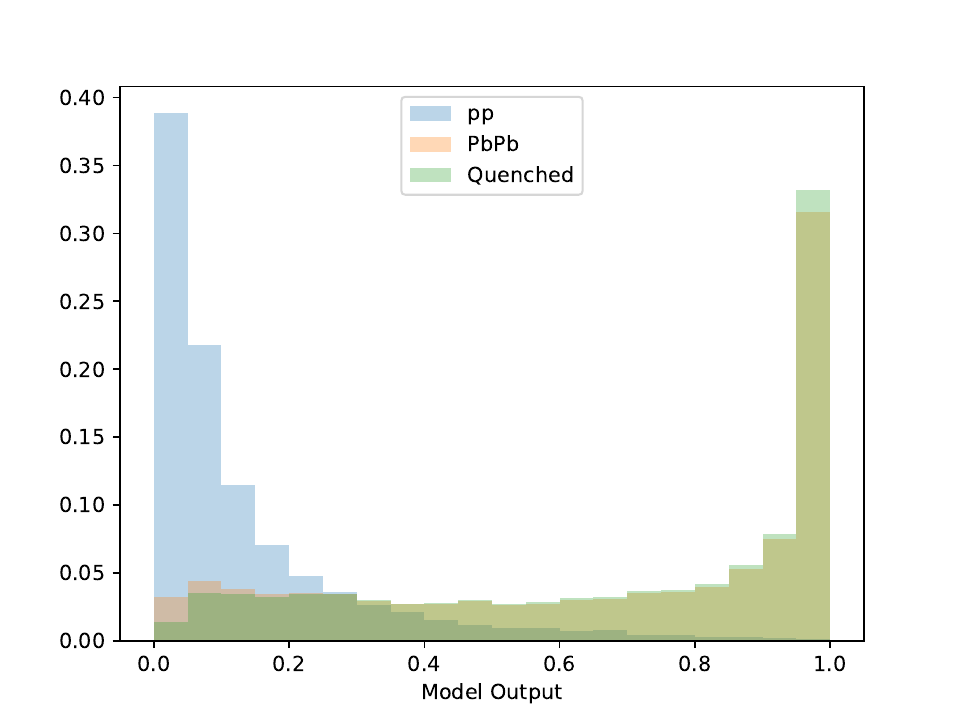}
        \caption{no UE, EFPs}
        \label{fig:dnn_out_efp_noue}
    \end{subfigure}
    \hfill
    \begin{subfigure}[b]{0.49\textwidth}
        \centering
        \includegraphics[width=\textwidth]{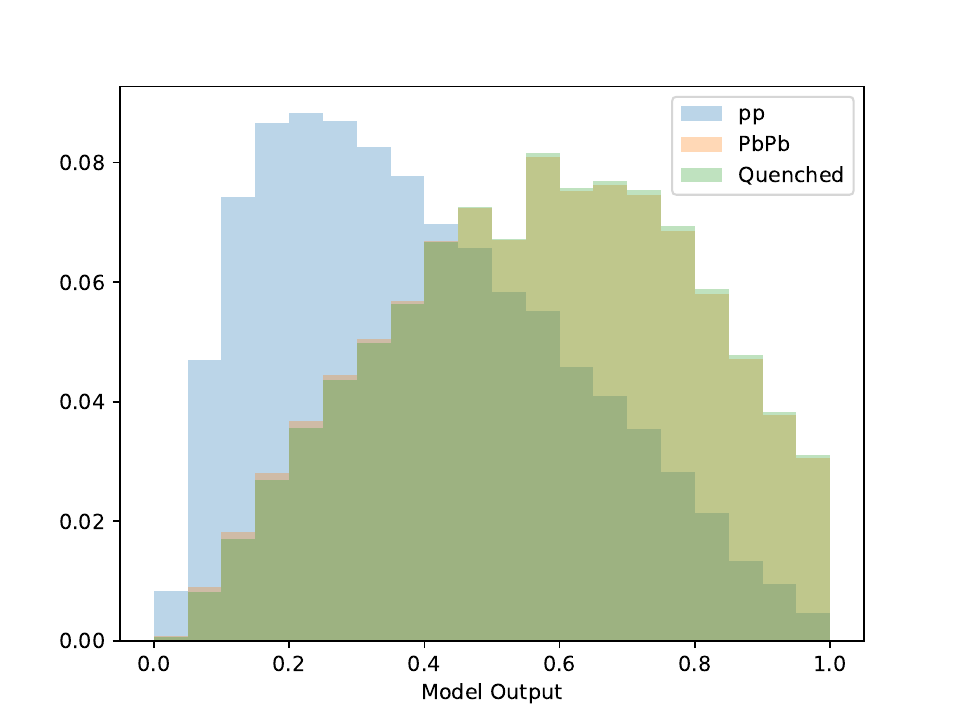}
        \caption{with UE, EFPs}
        \label{fig:dnn_out_efp_ue}
    \end{subfigure}

    \begin{subfigure}[b]{0.49\textwidth}
        \centering
        \includegraphics[width=\textwidth]{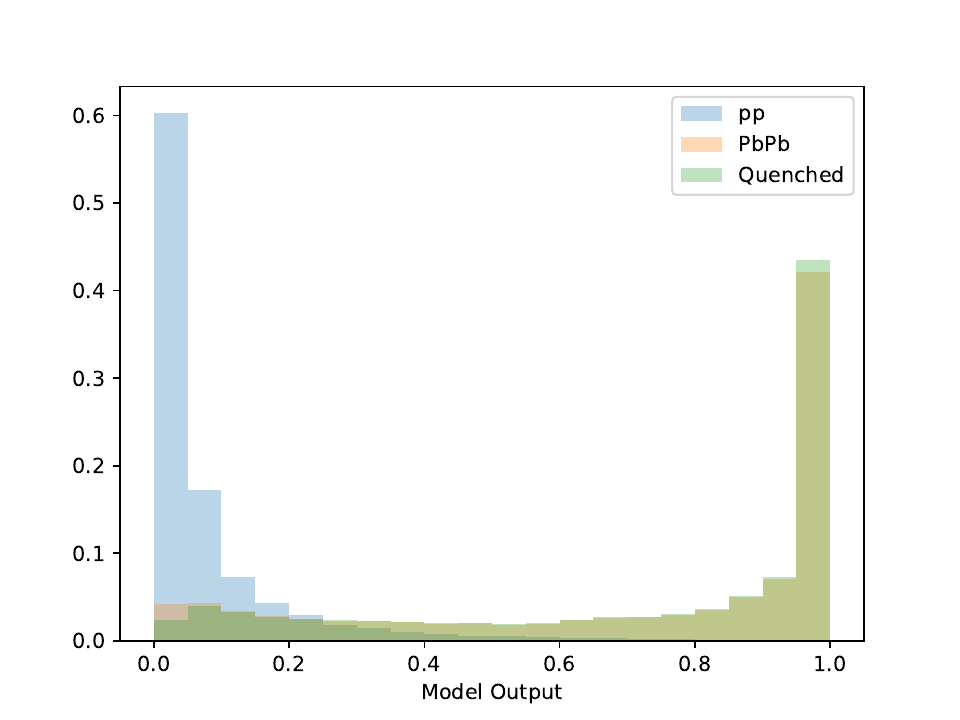}
        \caption{no UE, extended EFPs}
        \label{fig:dnn_out_efpext_noue}
    \end{subfigure}
    \hfill
    \begin{subfigure}[b]{0.49\textwidth}
        \centering
        \includegraphics[width=\textwidth]{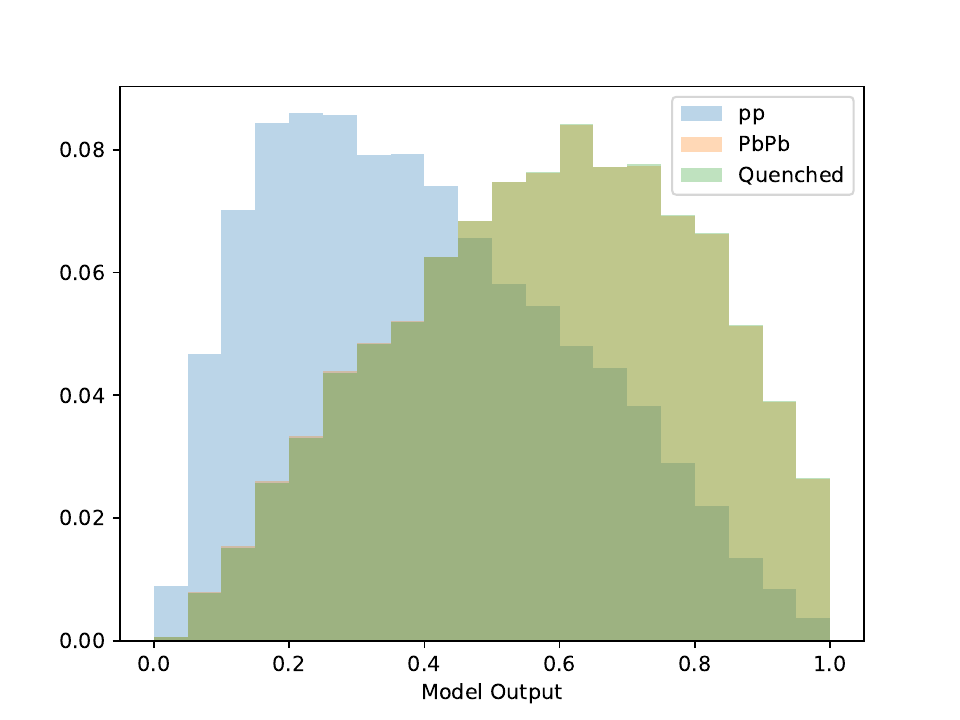}
        \caption{with UE, extended EFPs}
        \label{fig:dnn_out_efpext_ue}
    \end{subfigure}

    \caption{Classifier output score distributions for NN models across input sets and UE configurations.}
    \label{fig:dnn_output_six}
\end{figure}
\noindent  results presented in Ref.~\cite{ArrudaGoncalves:2025wtb}. Relative to the linear baselines in Fig.~\ref{fig:linear_roc_six}, the NN uplift is modest but systematic --- AUC gains are typically at the level of about $2$--$9\%$ in the no UE case and about $0.5$--$2\%$ with UE --- while preserving the hierarchy across bases. The EFP ROCs (Figs.~\ref{fig:dnn_roc_efp_noue},~\ref{fig:dnn_roc_efp_ue}) retain higher performance, when compared to the $N$-subjettiness curves, and the extended EFPs (Figs.~\ref{fig:dnn_roc_efpext_noue},~\ref{fig:dnn_roc_efpext_ue}) yield the best overall discrimination in both settings.

The class-conditional score distributions in Fig.~\ref{fig:dnn_output_six} mirror these trends. For $N$-subjettiness without UE (Fig.~\ref{fig:dnn_out_nsub_noue}), the pp and PbPb peaks are essentially disjoint with a clear valley, in marked contrast to the substantial overlap seen with LDA (cf.~Figs.~\ref{fig:out_nsub_noue},~\ref{fig:out_nsub_ue}). With UE (Fig.~\ref{fig:dnn_out_nsub_ue}), the separation reduces but remains visibly better than the linear case. EFPs (Figs.~\ref{fig:dnn_out_efp_noue},~\ref{fig:dnn_out_efp_ue}) show narrower overlaps than $N$-subjettiness under both conditions, and the extended EFPs (Figs.~\ref{fig:dnn_out_efpext_noue},~\ref{fig:dnn_out_efpext_ue}) exhibit the cleanest separation overall, fully consistent with the ROC ordering.

\subsection{Energy flow networks}
\label{sssec:efn}

\begin{figure}[!htbp]
    \centering
    \includegraphics[width=\textwidth]{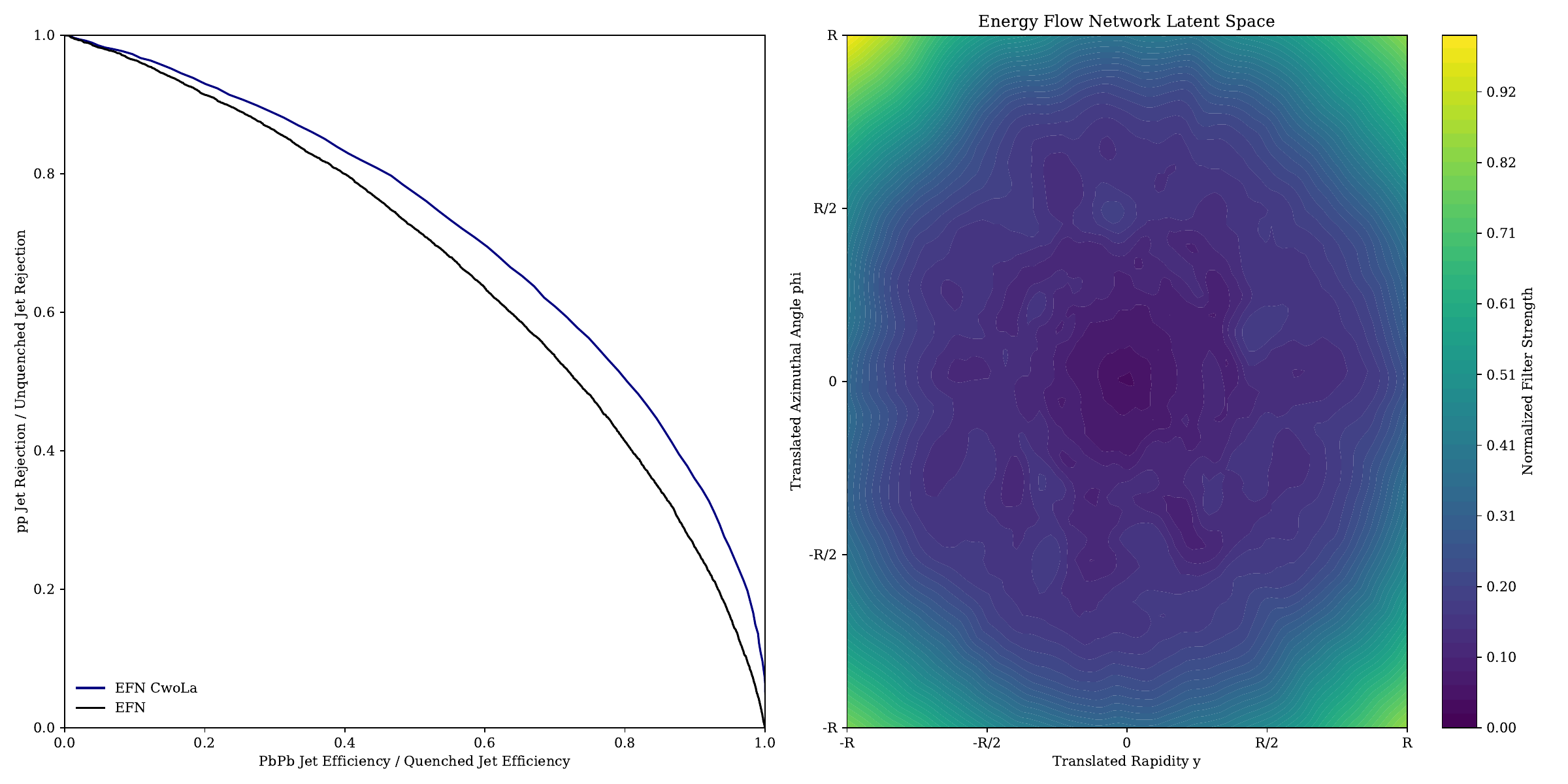}
    \caption{ROC curve for EFN (left) and latent space average visualization (right).}
    \label{fig:efn_roc}
\end{figure}

\noindent EFNs process sets of jet constituents with permutation invariance and transverse-momentum-weighted latent space, targeting IRC safety at the representation level. EFNs are evaluated with consistent preprocessing, cross-validation, and early stopping, and ROC AUC, latent-space structure, output distributions, learning curves, and cross-validation stability plots are reported.

\begin{figure}[!htbp]
    \centering
    \begin{subfigure}[b]{0.38\textwidth}
        \centering
        \includegraphics[width=\textwidth]{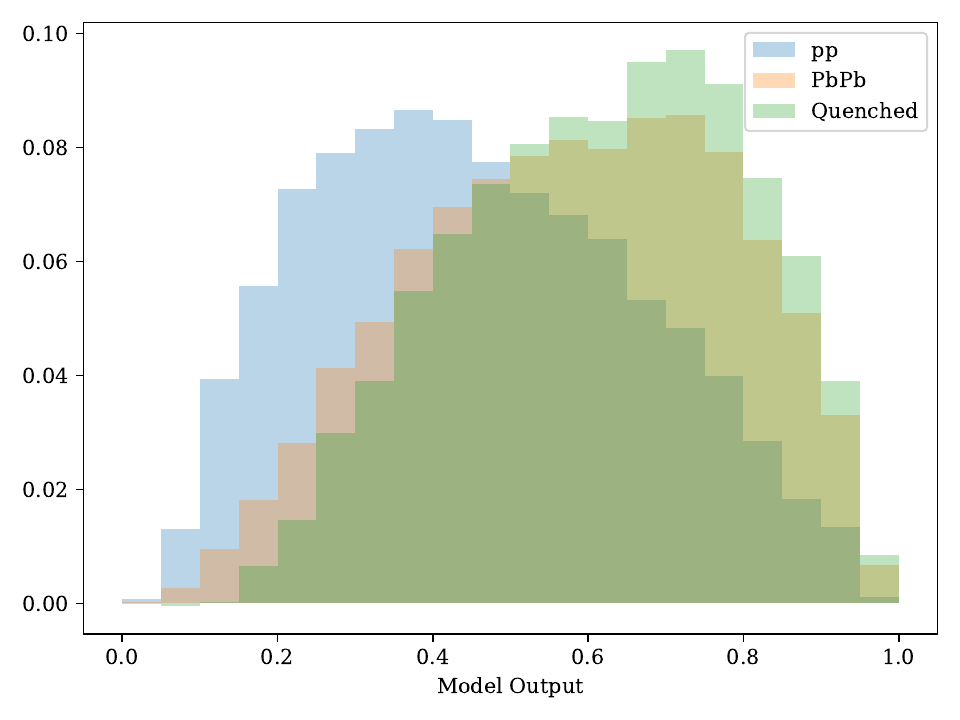}
        \caption{Output score distributions}
        \label{fig:efn_out}
    \end{subfigure}
    \hfill
    \begin{subfigure}[b]{0.58\textwidth}
        \centering
        \includegraphics[width=\textwidth]{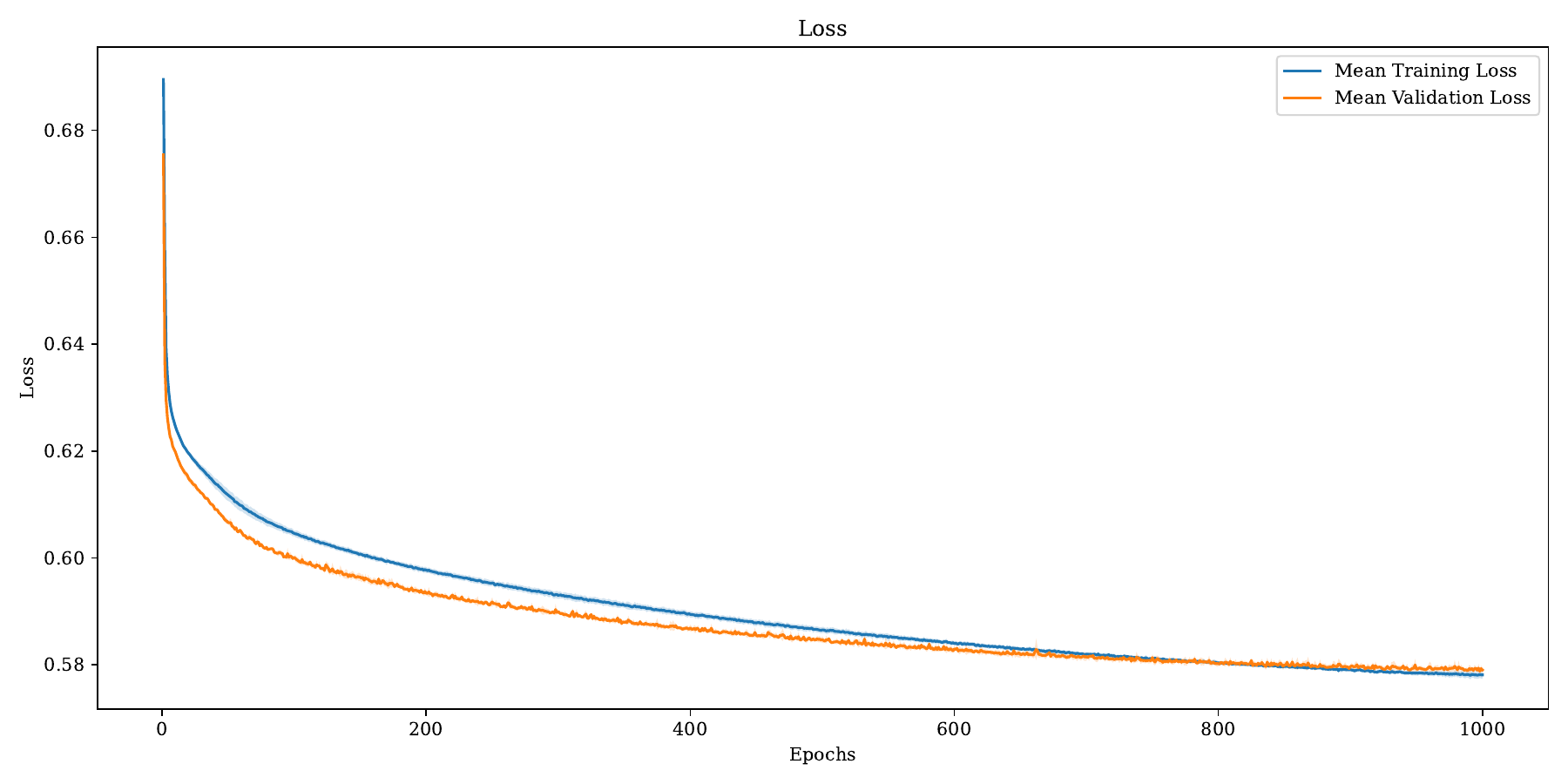}
        \caption{Training/validation loss}
        \label{fig:efn_loss}
    \end{subfigure}
    \caption{EFN output distributions and learning curve.}
    \label{fig:efn_out_loss}
\end{figure}
    
All EFN results in this subsection include UE contamination. The left panel of Fig.~\ref{fig:efn_roc} shows the supervised ROC together with the CWoLa ROC on the test set; CWoLa yields a small performance increase. The right panel of Fig.~\ref{fig:efn_roc} displays the average latent space response in the translated pseudorapidity-azimuth plane, revealing a radially symmetric structure, with fluctuations in the mid-range. 

\begin{figure}[!htbp]
    \centering
    \includegraphics[width=0.8\textwidth]{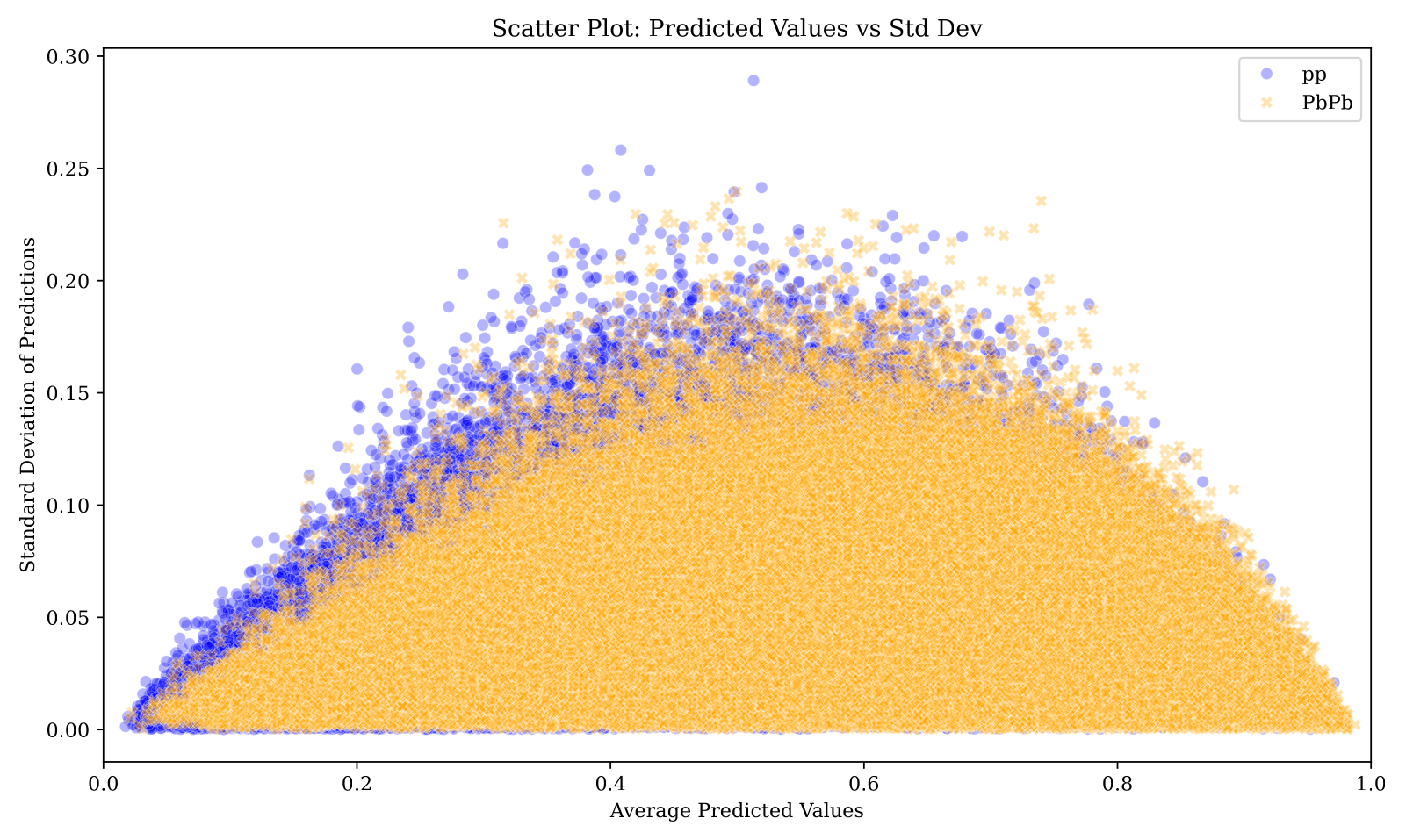}
    \caption{Cross-validation stability: AUC per fold (with mean and variance) or output mean vs.\ standard deviation.}
    \label{fig:efn_cv_stability}
\end{figure}

The class-conditional score densities in Fig.~\ref{fig:efn_out} exhibit some separation between pp and PbPb jets, with a more pronounced valley and slightly less overlap when compared to the observable baselines. Consistently, the EFN achieves an AUC improvement of about five percentage points relative to the best linear/non-linear baseline that also includes UE, confirming the benefit of learning directly from constituents within an IRC-aware architecture. The training and validation losses in Fig.~\ref{fig:efn_loss} decrease smoothly and plateau with a negligible gap; there are no signs of underfitting or overfitting, and early-stopping behavior is consistent with stable convergence.

To assess training stability across folds, Fig.~\ref{fig:efn_cv_stability} presents the scatter of the fold-averaged classifier output versus its standard deviation across folds: the taco plot. The dispersion is smallest near scores close to 0 or 1 and remains below about 0.25 through the central region, indicating limited fold-to-fold variability. This level of stability and performance sets a new baseline for the observable-augmented variants considered next.

\subsection{Observable-enhanced EFNs}
\label{sssec:oefn}

While EFNs encode the full set of jet constituents in an IRC-safe way, they do not directly incorporate high-level observables known to capture important aspects of quenching. Observable-enhanced EFNs (oEFNs) address this by concatenating global observables such as $N$-subjettiness or EFPs to the latent representation learned by the EFN. This hybrid approach allows the model to benefit from explicit, physics-motivated features, while keeping permutation invariance if the features also have this property. This subsection gauges the gains from such enhancements.

\begin{figure}[!htbp]
    \centering
    \includegraphics[width=.65\textwidth]{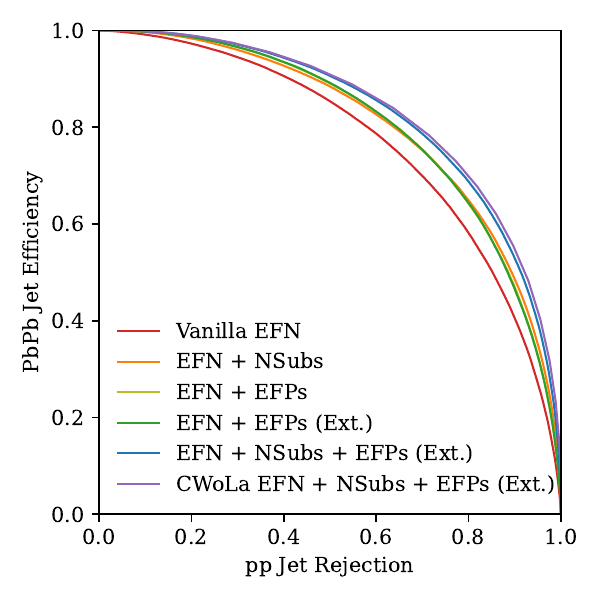}
    \caption{ROC curves for oEFN classifiers under different observable augmentations, with and without UE contamination.}
    \label{fig:oefn_roc}
\end{figure}

The aggregated ROC comparison in Fig.~\ref{fig:oefn_roc} shows a monotonic improvement as observables are concatenated to the EFN latent space. Relative to the vanilla EFN (AUC $\simeq 0.765$), extending the latent space with either $N$-subjettiness or EFPs lifts the AUC to about $0.805$, using extended EFPs yields about $0.810$, combining $N$-subjettiness with standard EFPs reaches $0.821$, and combining $N$-subjettiness with extended EFPs attains about $0.827$. It is worth underlining that this is under operation in the UE-contaminated setting with MR, and hence these results quantify real gains in a realistic regime.

\begin{figure}[!htbp]
    \centering
    \begin{subfigure}[b]{0.48\textwidth}
        \centering
        \includegraphics[width=\textwidth]{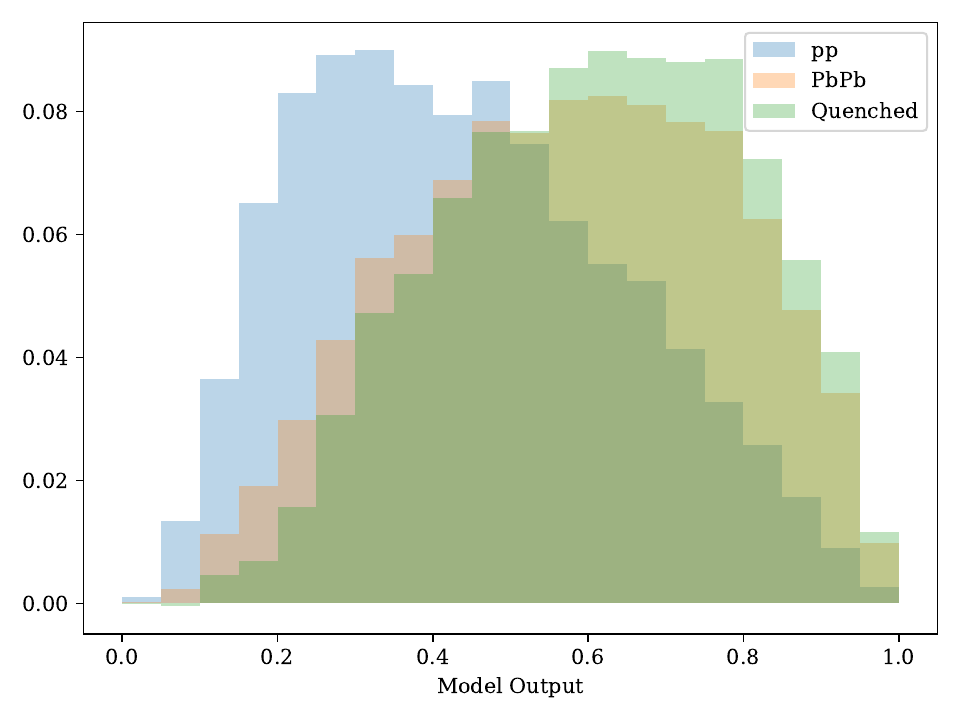}
        \caption{Vanilla}
        \label{fig:oefn_out_vanilla}
    \end{subfigure}
    \hfill
    \begin{subfigure}[b]{0.48\textwidth}
        \centering
        \includegraphics[width=\textwidth]{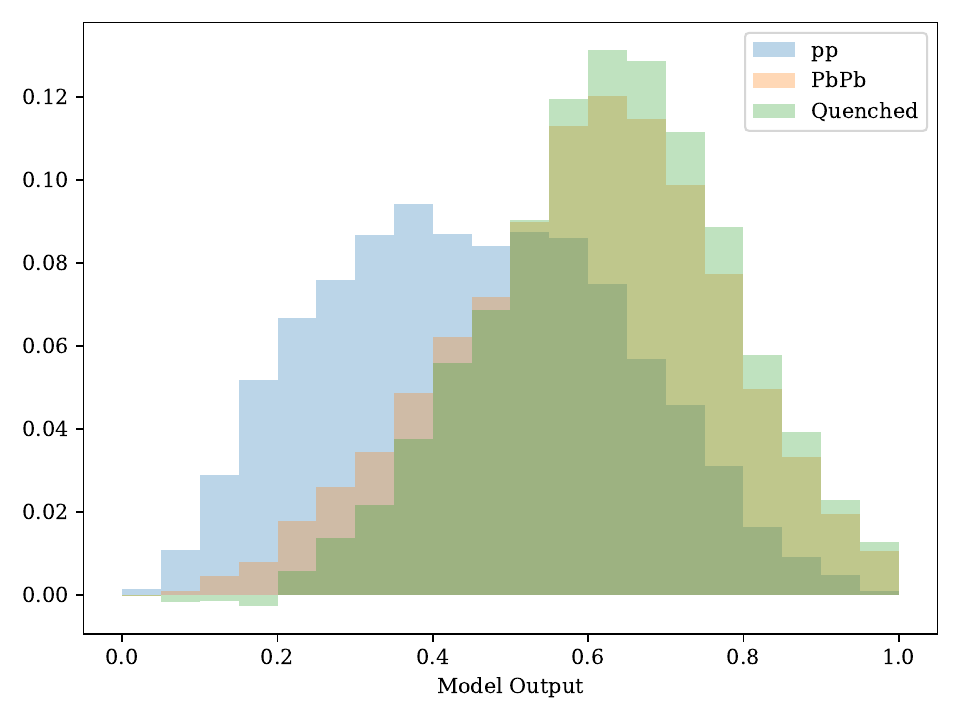}
        \caption{$N$-subjettiness}
        \label{fig:oefn_out_nsub}
    \end{subfigure}
    \\
    \begin{subfigure}[b]{0.48\textwidth}
        \centering
        \includegraphics[width=\textwidth]{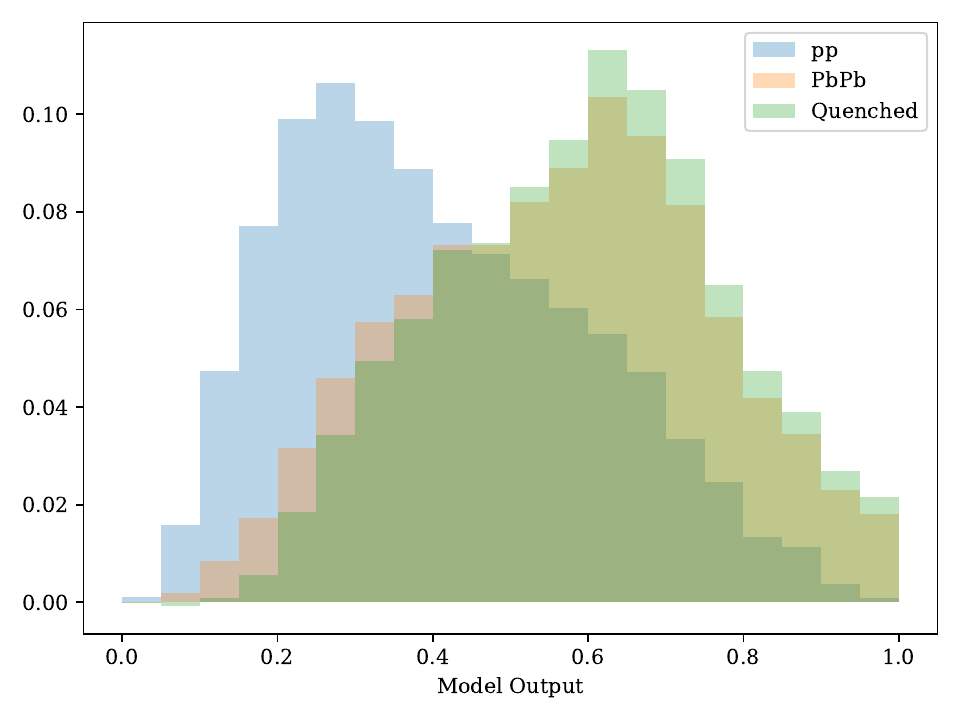}
        \caption{EFPs}
        \label{fig:oefn_out_efp}
    \end{subfigure}
    \hfill
    \begin{subfigure}[b]{0.48\textwidth}
        \centering
        \includegraphics[width=\textwidth]{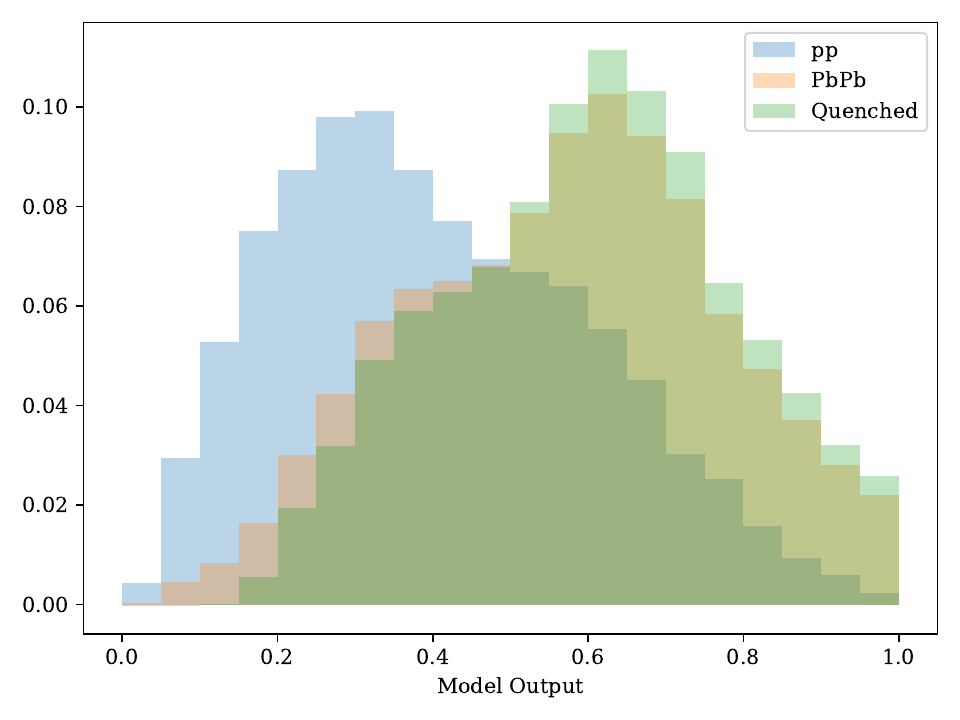}
        \caption{Extended EFPs}
        \label{fig:oefn_out_efpext}
    \end{subfigure}
    \\
    \begin{subfigure}[b]{0.48\textwidth}
        \centering
        \includegraphics[width=\textwidth]{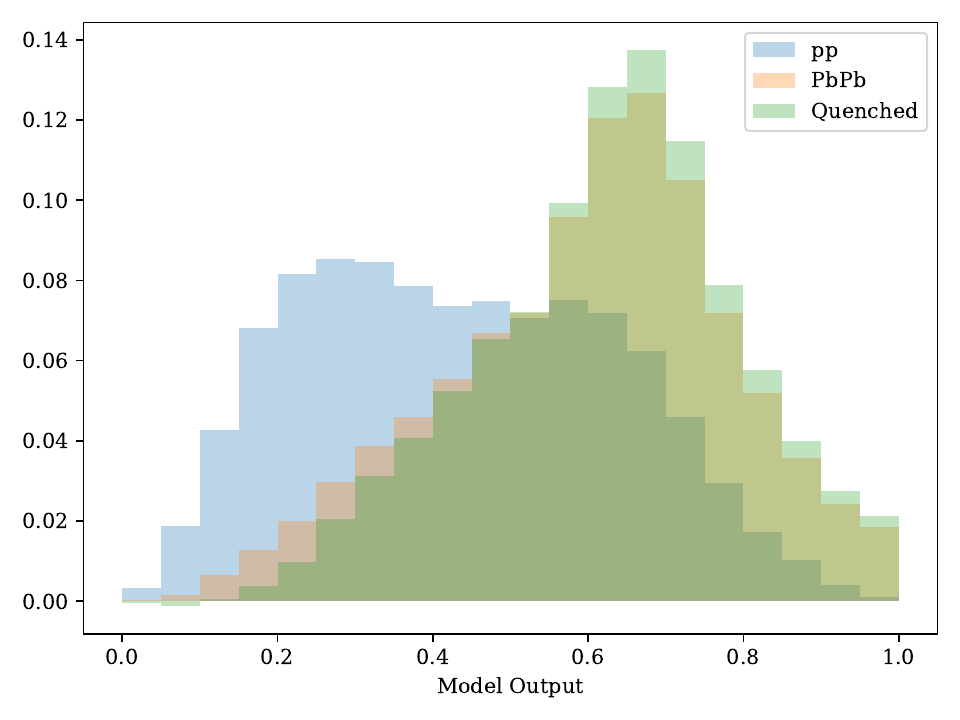}
        \caption{$N$-subjettiness + EFPs}
        \label{fig:oefn_out_both}
    \end{subfigure}
    \hfill
    \begin{subfigure}[b]{0.48\textwidth}
        \centering
        \includegraphics[width=\textwidth]{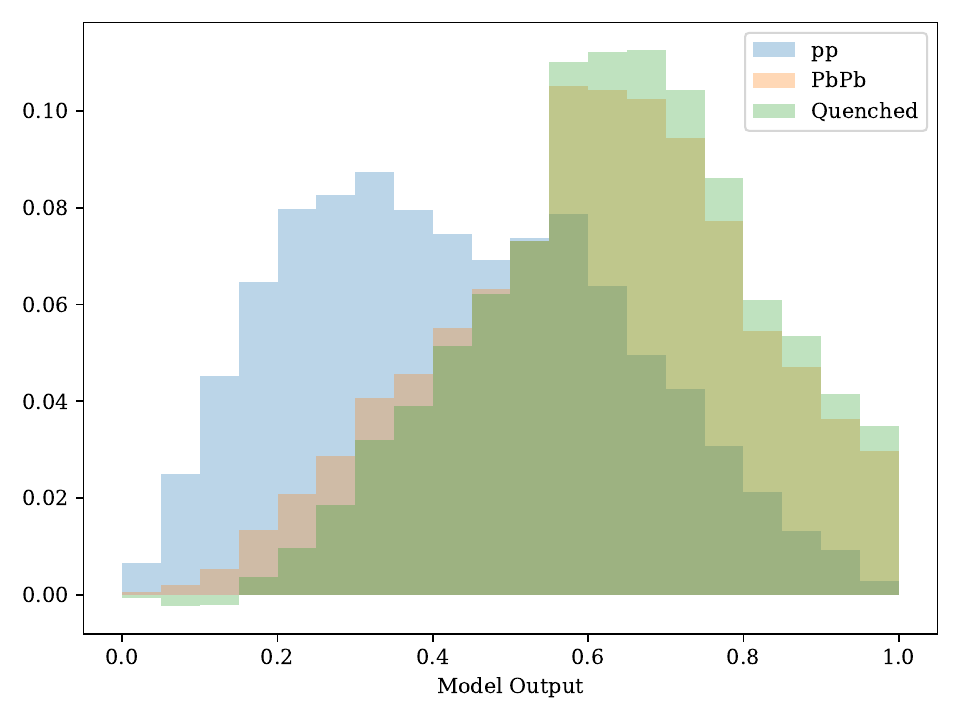}
        \caption{$N$-subjettiness + Extended EFPs}
        \label{fig:oefn_out_both_ext}
    \end{subfigure}
    \caption{Classifier output score distributions for oEFNs under different input augmentations.}
    \label{fig:oefn_outputs}
\end{figure}

The class-conditional score densities in Fig.~\ref{fig:oefn_outputs} reflect the same ordering. Comparing panels~\ref{fig:oefn_out_vanilla},~\ref{fig:oefn_out_nsub}, and~\ref{fig:oefn_out_efp}, the pp and PbPb peaks separate progressively as the AUC rises, with a slightly wider valley. The extended EFP augmentation (Fig.~\ref{fig:oefn_out_efpext}) adds a further, smaller improvement. The strongest separation appears when combining observables (Figs.~\ref{fig:oefn_out_both} and~\ref{fig:oefn_out_both_ext}); in the best case the PbPb distribution is substantially shifted to the right relative to the vanilla configuration, consistent with the ROC uplift and improved background rejection.

\subsection{Moment EFNs}
\label{sssec:mefn}

\begin{figure}[!htbp]
    \centering
    \includegraphics[scale=.4]{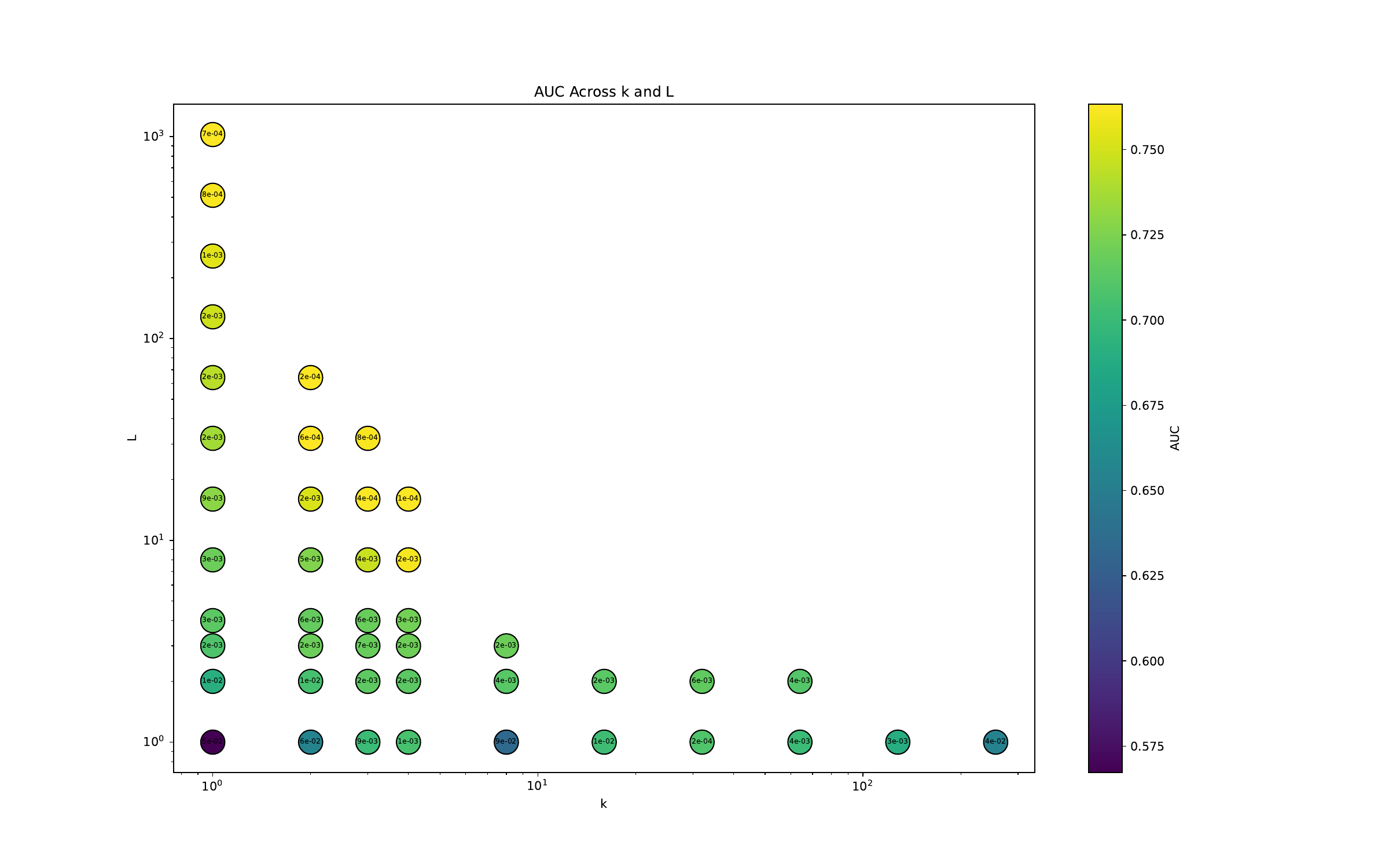}
    \caption{Two-dimensional map of AUC values across latent space dimension $L$ and MEFN order $k$.}
    \label{fig:mefn_auc_heatmap}
\end{figure}

Moment energy flow networks (MEFNs) extend EFNs by passing higher-order moments of the learned latent coordinates to the classifier (an operation known as moment pooling), enabling EFN-level discrimination with far smaller latent space size. A scan across order $k$ and latent dimension $L$ under UE contamination is presented and summarizes the landscape in a two-dimensional AUC map (Fig.~\ref{fig:mefn_auc_heatmap}). The grid exhibits clear structure across $(k,L)$ rather than random variation, delineating regions where compact models already approach the best observed performance.

\begin{figure}[!htbp]
    \centering
    \includegraphics[width=1\textwidth]{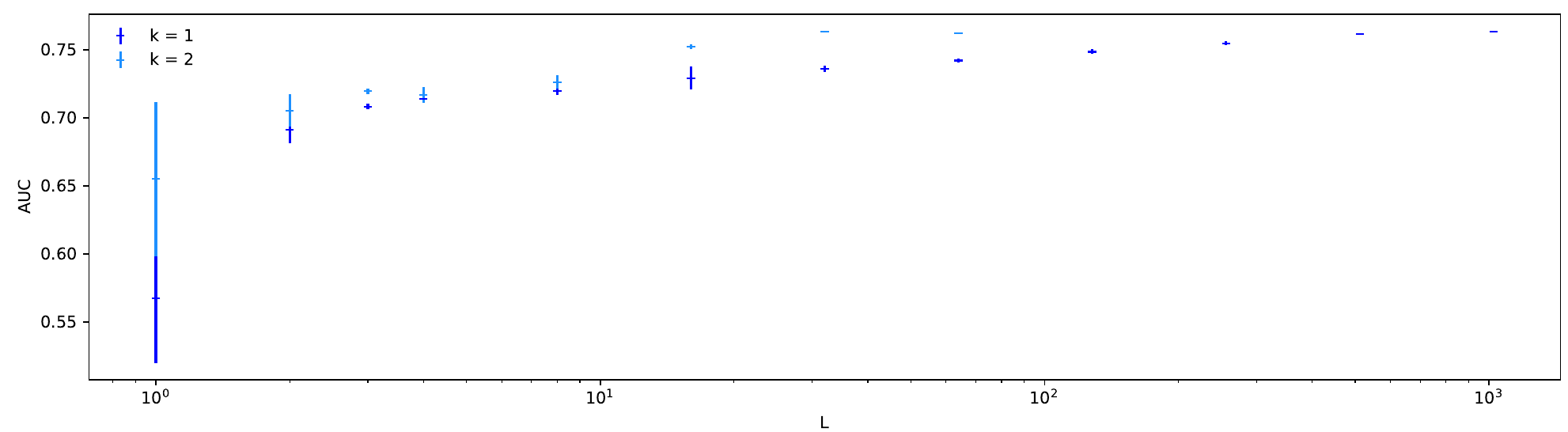}
    \\
        \includegraphics[width=1.005\textwidth]{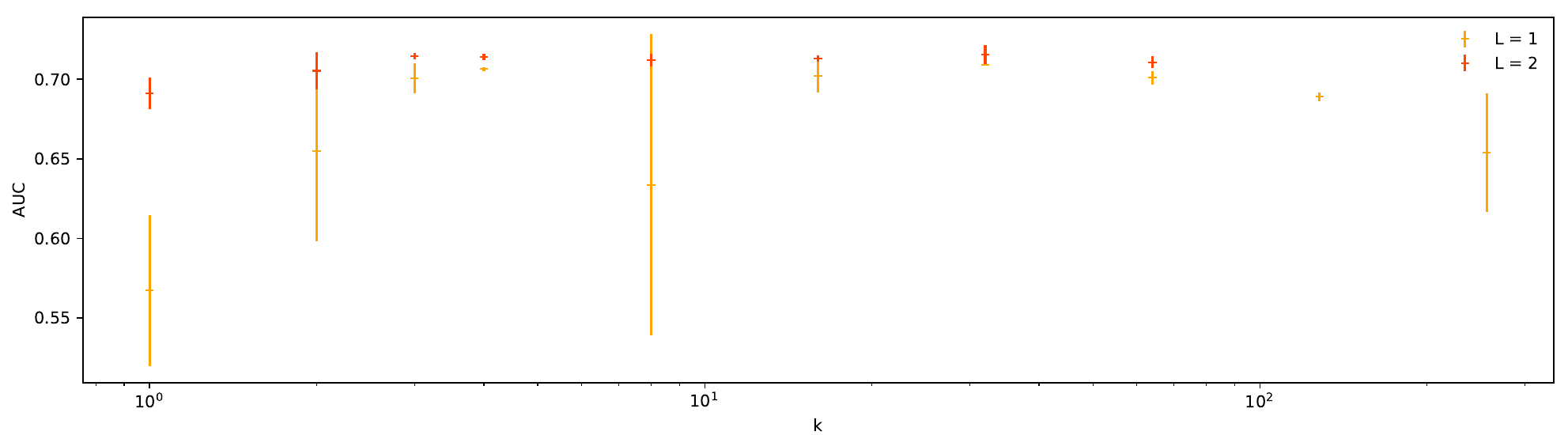}
    \caption{Evolution of $1 - \mathrm{AUC}$ with latent space dimension (top) and MEFN order $k$ (bottom).}
    \label{fig:mefn_one_minus_auc}
\end{figure}

Scaling trends are explicit in the one-dimensional sweeps. At fixed order, increasing the latent size improves AUC up to a visible plateau; conversely, at fixed latent size, raising the order yields gains that taper as $k$ grows (Figs.~\ref{fig:mefn_one_minus_auc}). These curves make the diminishing-returns trade-off transparent and guide the choice of a compact working point. 

\begin{figure}[!htbp]
    \centering
    \includegraphics[width=\textwidth]{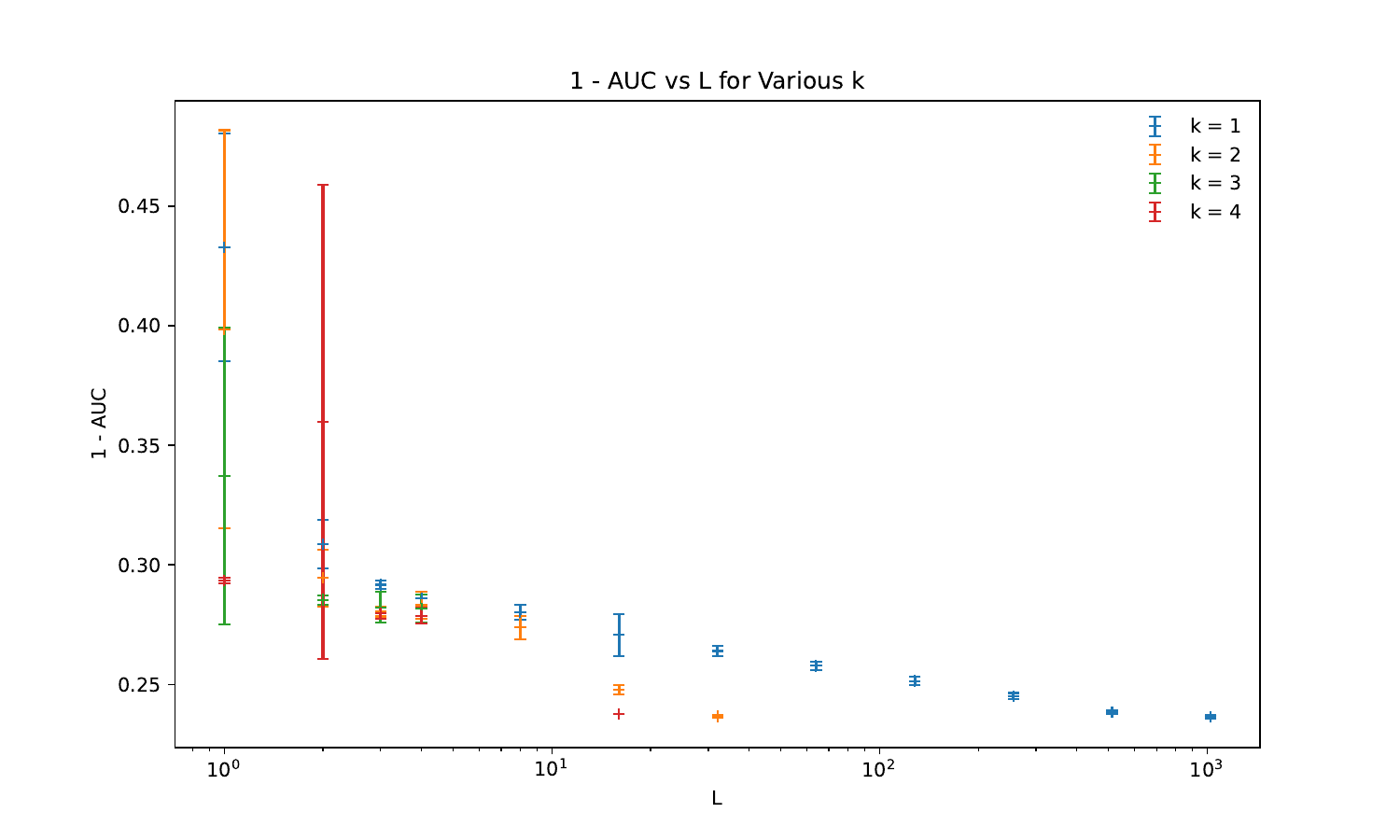}
    \\
    \includegraphics[width=\textwidth]{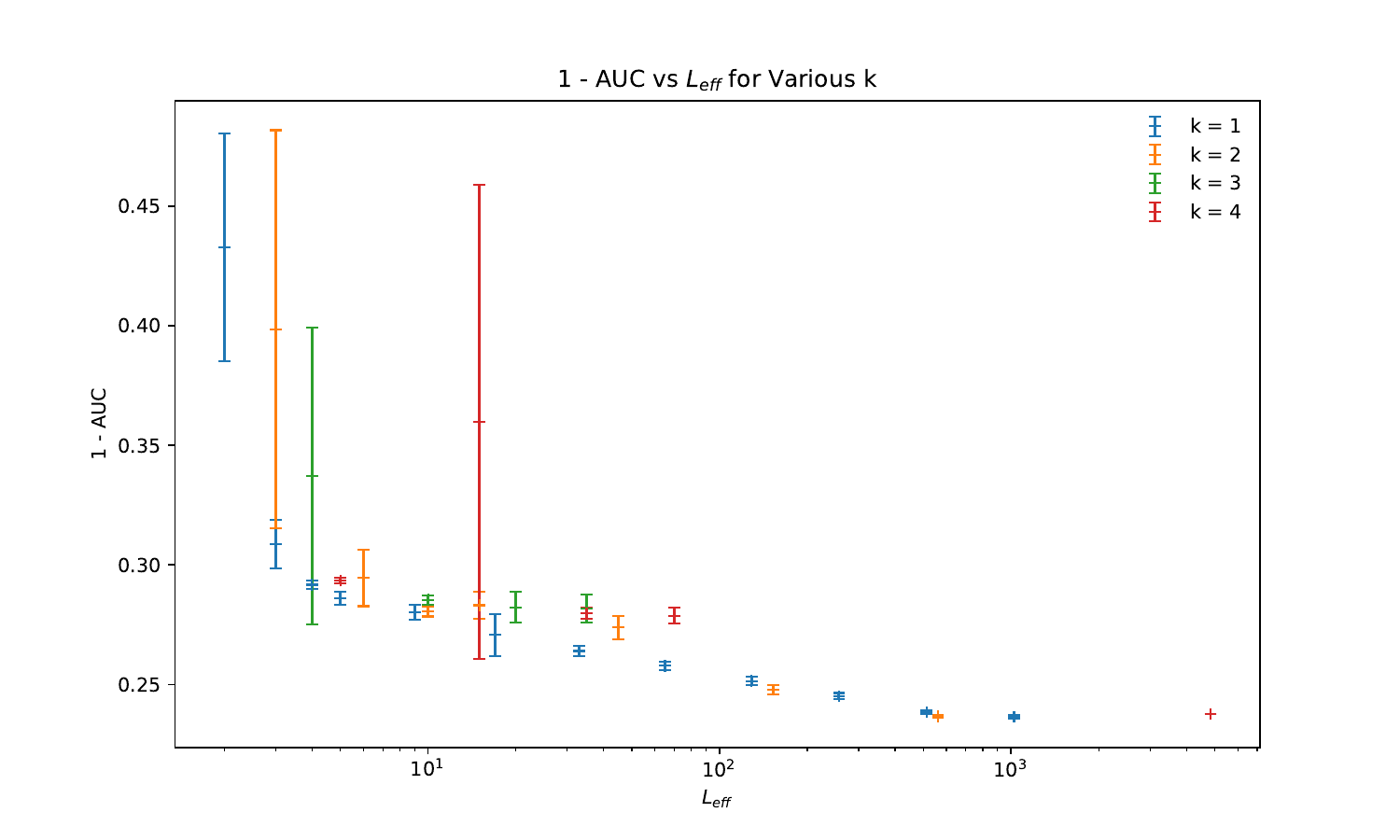}
    \caption{$1 - \mathrm{AUC}$ vs.\ latent space dimension (top) and effective latent dimension (bottom).}
    \label{fig:mefn_one_minus_auc_eff}
\end{figure}

\begin{figure}[!htbp]
    \centering
    \includegraphics[width=\textwidth]{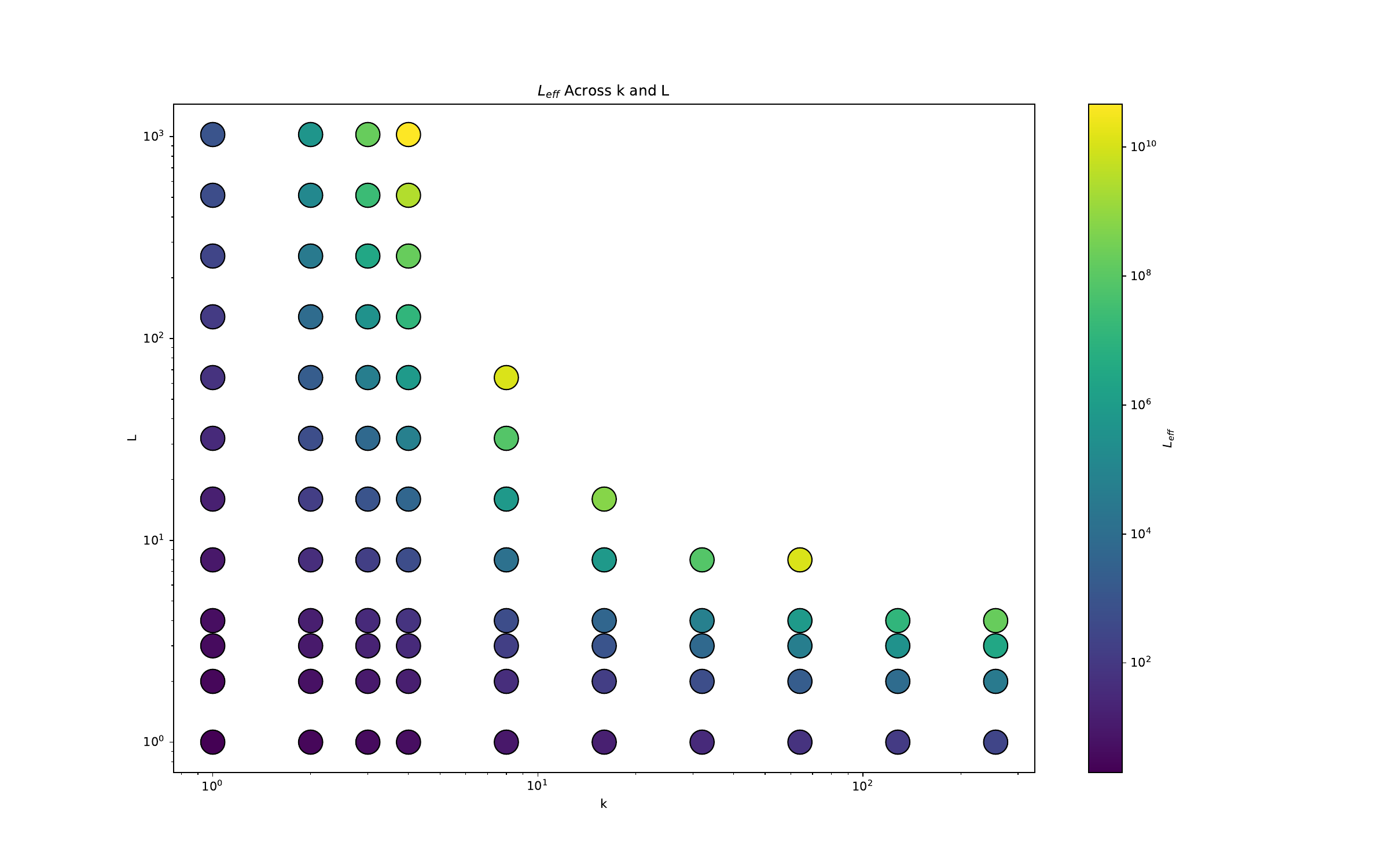}
    \\
    \includegraphics[width=\textwidth]{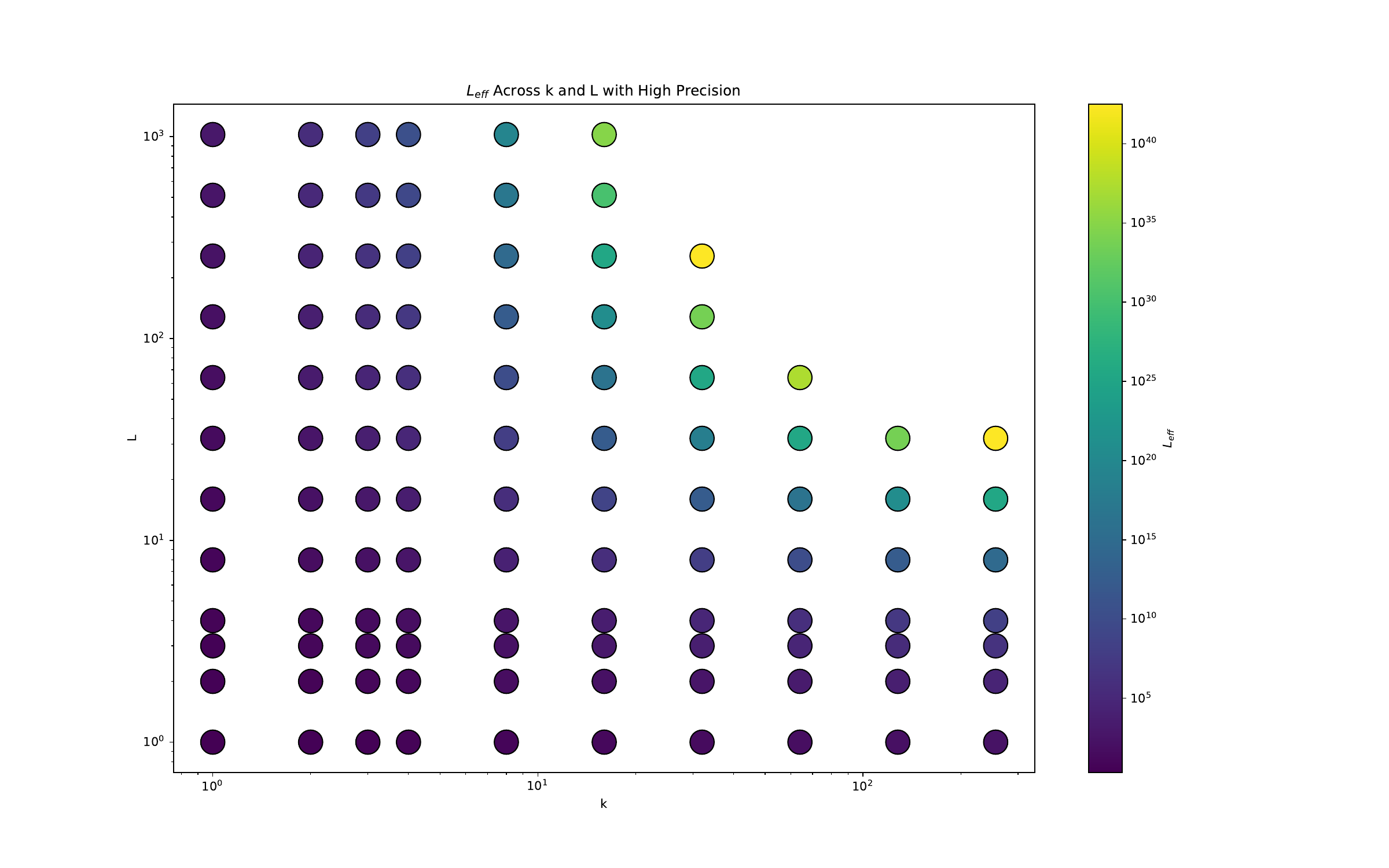}
    \caption{Effective latent space dimension across MEFN order $k$ (top) and latent space dimension $L$ (bottom).}
    \label{fig:mefn_eff_latent}
\end{figure}

Re-expressing performance against the effective latent dimension $L_{\mathrm{eff}}$ reduces the spread across configurations relative to plotting against $L$ alone (Fig.~\ref{fig:mefn_one_minus_auc_eff}). The corresponding maps of $L_{\mathrm{eff}}$ over $(k,L)$ (Fig.~\ref{fig:mefn_eff_latent}) further show how higher orders with large latent space dimension \(L\) quickly scale \(L_{\mathrm{eff}}\).

In practice, this yields efficient choices that retain EFN-level AUC with markedly fewer channels (as reflected in the scan table): representative points such as $k=2,\,L=32$ and $k=3,\,L=16$ lie near the efficient frontier and match baseline EFN performance within uncertainties while reducing model size and enhancing interpretability. The AUC for the complete set of runs is consolidated in Tab.~\ref{tab:mefn_scan} alongside a similar table for the observable-enhanced version, Tab.~\ref{tab:omefn_scan}.

\subsection{Overall comparison}
\label{ssec:overall_comp}

Tables~\ref{tab:model_comp_noUE}--\ref{tab:model_comp_UE} consolidate the performance of all tested models in terms of ROC AUC, except for the moment variants. Results without UE contamination are presented in Tab.~\ref{tab:model_comp_noUE}, while results with UE contamination are presented in Tab.~\ref{tab:model_comp_UE}. In the realistic setting, EFNs improve substantially over observable-only baselines, and oEFNs provide the best overall AUCs, reaching $\simeq0.83$. A detailed MEFN/oMEFN scan is presented in Tabs.~\ref{tab:mefn_scan}--\ref{tab:omefn_scan} and shows many compact configurations achieving oEFN-level performance, clarifying the trade-off between interpretability/compactness and raw capacity.

\begin{table}[!htbp]
\centering
\small
\begin{tabular}{lc}
\toprule
Model & AUC \\
\midrule
LDA EFP          & 0.8675 \\
LDA EFP Ext.     & 0.9234 \\
LDA $N$-subjettiness         & 0.8314 \\
NN EFP          & 0.9067 \\
NN EFP Ext.     & \textbf{0.9420} \\
NN $N$-subjettiness         & 0.9232 \\
\bottomrule
\end{tabular}
\caption{Model comparison (supervised) without UE contamination.}
\label{tab:model_comp_noUE}
\end{table}

\begin{table}[!htbp]
\centering
\small
\begin{tabular}{lc}
\toprule
Model & AUC \\
\midrule
LDA EFP                 & 0.6964 \\
LDA EFP Ext.            & 0.7132 \\
LDA $N$-subjettiness    & 0.6900 \\
NN EFP                 & 0.7142 \\
NN EFP Ext.            & 0.7176 \\
NN $N$-subjettiness    & 0.7075 \\
EFN                     & 0.7651 $\pm$ 0.0004 \\
EFN + EFP               & 0.8050 $\pm$ 0.0002 \\
EFN + EFP Ext.          & 0.8104 $\pm$ 0.0004 \\
EFN + $N$-subjettiness  & 0.8053 $\pm$ 0.0003 \\
EFN + EFP + $N$-subjettiness        & 0.8206 $\pm$ 0.0001 \\
EFN + EFP Ext. + $N$-subjettiness   & \textbf{0.8265 $\pm$ 0.0011} \\
\bottomrule
\end{tabular}
\caption{Model comparison (supervised) with UE contamination. Uncertainties denote the cross-validation standard deviation across all folds.}
\label{tab:model_comp_UE}
\end{table}
\section{Conclusion}
\label{sec:concl}

A per-jet discrimination study between pp and PbPb jets in a realistic (UE+MR) setting, combining physics-motivated observables with constituent-level, IRC-aware representations is presented. Linear and shallow non-linear baselines on $N$-subjettiness and (extended) EFPs establish a clear hierarchy, with extended EFPs performing best among high-level observable-only models. EFNs trained on constituents deliver a marked uplift over these baselines. Augmenting EFNs with standardized high-level observables (oEFNs) yields further, robust gains, achieving ROC AUC $\simeq 0.83$ with cross-validation stability. 

Moment EFNs (MEFNs), which make use of moment pooling (defined in Eq.~\ref{eq:mefn}), have been trained and shown to match EFN-level AUC in the jet quenching problem. Their observable-enhanced counterparts (oMEFNs) reach oEFN-level performance across a broad $(k,L)$ band, providing a practical route to smaller and more interpretable latent spaces without sacrificing accuracy.

Several aspects are natural targets for future work. On the physics side: centrality, jet-$p_T$ and jet-$R$ dependence would further complete the study for the usability of the approach across these observables. More importantly, studies across different physics models would allow a Pareto Frontier study of the jet quenching problem, such as in Ref.~\cite{Gambhir:2025xim}. On the learning side: comparisons to alternative set/graph architectures (e.g.\ ParticleNet/transformers), calibration and uncertainty estimation, and systematic robustness to UE modeling and subtraction choices. Finally, applying (o)EFN/(o)MEFN weak supervision directly to data, in the apples-to-apples setup, would close the loop to experimental deployment~\cite{ArrudaGoncalves:2025wtb}.

Taken together, these results establish EFN-based approaches --- especially oEFNs and oMEFNs --- as practical, robust tools for jet-quenching classification in realistic UE+MR environments. The main message of this paper is that combining constituent-level information with high-level observables is the best approach to the jet quenching problem: ``all is more''. Crucially, by operating on constituents they capture non-linear, multi-particle structure that escapes simple correlations among a handful of high-level observables, and they do so in a compact and relatively more interpretable form~\cite{ArrudaGoncalves:2025wtb}.

\begin{table}[H]
\centering
\small

\begin{subtable}[t]{0.48\linewidth}
\centering
\begin{tabular}{rrc}
\toprule
$k$ & $L$ & AUC $\pm$ CV std. \\
\midrule
1   & 1024 & $0.7634 \pm 0.0007$ \\
2   & 32   & $0.7632 \pm 0.0006$ \\
3   & 16   & $0.7624 \pm 0.0004$ \\
2   & 64   & $0.7623 \pm 0.0002$ \\
4   & 16   & $0.7623 \pm 0.0001$ \\
1   & 512  & $0.7615 \pm 0.0008$ \\
4   & 8    & $0.7608 \pm 0.0016$ \\
1   & 256  & $0.7548 \pm 0.0013$ \\
2   & 16   & $0.7521 \pm 0.0018$ \\
1   & 128  & $0.7485 \pm 0.0017$ \\
1   & 64   & $0.7421 \pm 0.0017$ \\
1   & 32   & $0.7359 \pm 0.0022$ \\
1   & 16   & $0.7293 \pm 0.0086$ \\
2   & 8    & $0.7262 \pm 0.0050$ \\
4   & 4    & $0.7212 \pm 0.0033$ \\
4   & 3    & $0.7201 \pm 0.0025$ \\
1   & 8    & $0.7196 \pm 0.0031$ \\
8   & 3    & $0.7193 \pm 0.0020$ \\
3   & 3    & $0.7177 \pm 0.0065$ \\
3   & 2    & $0.7146 \pm 0.0020$ \\
4   & 2    & $0.7142 \pm 0.0020$ \\
16  & 2    & $0.7131 \pm 0.0020$ \\
8   & 2    & $0.7121 \pm 0.0039$ \\
64  & 2    & $0.7108 \pm 0.0038$ \\
32  & 1    & $0.7092 \pm 0.0002$ \\
1   & 3    & $0.7083 \pm 0.0019$ \\
4   & 1    & $0.7066 \pm 0.0013$ \\
2   & 2    & $0.7054 \pm 0.0119$ \\
16  & 1    & $0.7021 \pm 0.0105$ \\
64  & 1    & $0.7009 \pm 0.0042$ \\
3   & 1    & $0.7006 \pm 0.0094$ \\
1   & 2    & $0.6912 \pm 0.0101$ \\
128 & 1    & $0.6891 \pm 0.0028$ \\
2   & 1    & $0.6549 \pm 0.0567$ \\
256 & 1    & $0.6537 \pm 0.0373$ \\
8   & 1    & $0.6337 \pm 0.0948$ \\
1   & 1    & $0.5673 \pm 0.0476$ \\
\bottomrule
\end{tabular}
\caption{MEFN results}
\label{tab:mefn_scan}
\end{subtable}
\hfill
\begin{subtable}[t]{0.48\linewidth}
\centering
\begin{tabular}{rrc}
\toprule
$k$ & $L$ & AUC $\pm$ CV std. \\
\midrule
64  & 1    & $0.8277 \pm 0.0005$ \\
64  & 2    & $0.8277 \pm 0.0004$ \\
16  & 1    & $0.8277 \pm 0.0005$ \\
32  & 2    & $0.8276 \pm 0.0005$ \\
3   & 1    & $0.8276 \pm 0.0006$ \\
128 & 1    & $0.8275 \pm 0.0007$ \\
1   & 2    & $0.8275 \pm 0.0006$ \\
1   & 3    & $0.8275 \pm 0.0005$ \\
1   & 4    & $0.8275 \pm 0.0008$ \\
256 & 1    & $0.8275 \pm 0.0006$ \\
2   & 2    & $0.8275 \pm 0.0007$ \\
3   & 2    & $0.8275 \pm 0.0004$ \\
32  & 1    & $0.8275 \pm 0.0005$ \\
4   & 2    & $0.8275 \pm 0.0008$ \\
2   & 1    & $0.8274 \pm 0.0008$ \\
8   & 1    & $0.8274 \pm 0.0008$ \\
16  & 2    & $0.8273 \pm 0.0005$ \\
1   & 1    & $0.8273 \pm 0.0006$ \\
1   & 32   & $0.8273 \pm 0.0007$ \\
1   & 8    & $0.8273 \pm 0.0008$ \\
2   & 3    & $0.8273 \pm 0.0007$ \\
2   & 8    & $0.8273 \pm 0.0007$ \\
3   & 3    & $0.8273 \pm 0.0009$ \\
1   & 128  & $0.8272 \pm 0.0005$ \\
2   & 16   & $0.8272 \pm 0.0007$ \\
2   & 4    & $0.8272 \pm 0.0007$ \\
8   & 2    & $0.8272 \pm 0.0011$ \\
4   & 3    & $0.8272 \pm 0.0010$ \\
1   & 16   & $0.8271 \pm 0.0012$ \\
2   & 32   & $0.8271 \pm 0.0006$ \\
3   & 4    & $0.8271 \pm 0.0010$ \\
1   & 256  & $0.8270 \pm 0.0007$ \\
1   & 64   & $0.8270 \pm 0.0010$ \\
4   & 4    & $0.8270 \pm 0.0010$ \\
4   & 1    & $0.8269 \pm 0.0009$ \\
2   & 64   & $0.8268 \pm 0.0008$ \\
1   & 512  & $0.8267 \pm 0.0009$ \\
1   & 1024 & $0.8265 \pm 0.0013$ \\
\bottomrule
\end{tabular}
\caption{oMEFN results}
\label{tab:omefn_scan}
\end{subtable}

\caption{AUC across $(k,L)$ settings for MEFN and oMEFN. Uncertainties denote the cross-validation standard deviation across folds.}
\label{tab:mefn_side_by_side}
\end{table}

\newpage
\acknowledgments
The author thanks José Guilherme Milhano and André Cordeiro for helpful discussions and for support by European Research Council under project ERC-2018-ADG-835105 YoctoLHC, as well as by FCT under contract PRT/BD/151554/2021. 

\appendix
\section{Stability Plots}
\label{app:loss}

\begin{figure}[!htb]
    \centering
    \begin{subfigure}[b]{0.49\textwidth}
        \centering
        \includegraphics[width=1.1\textwidth]{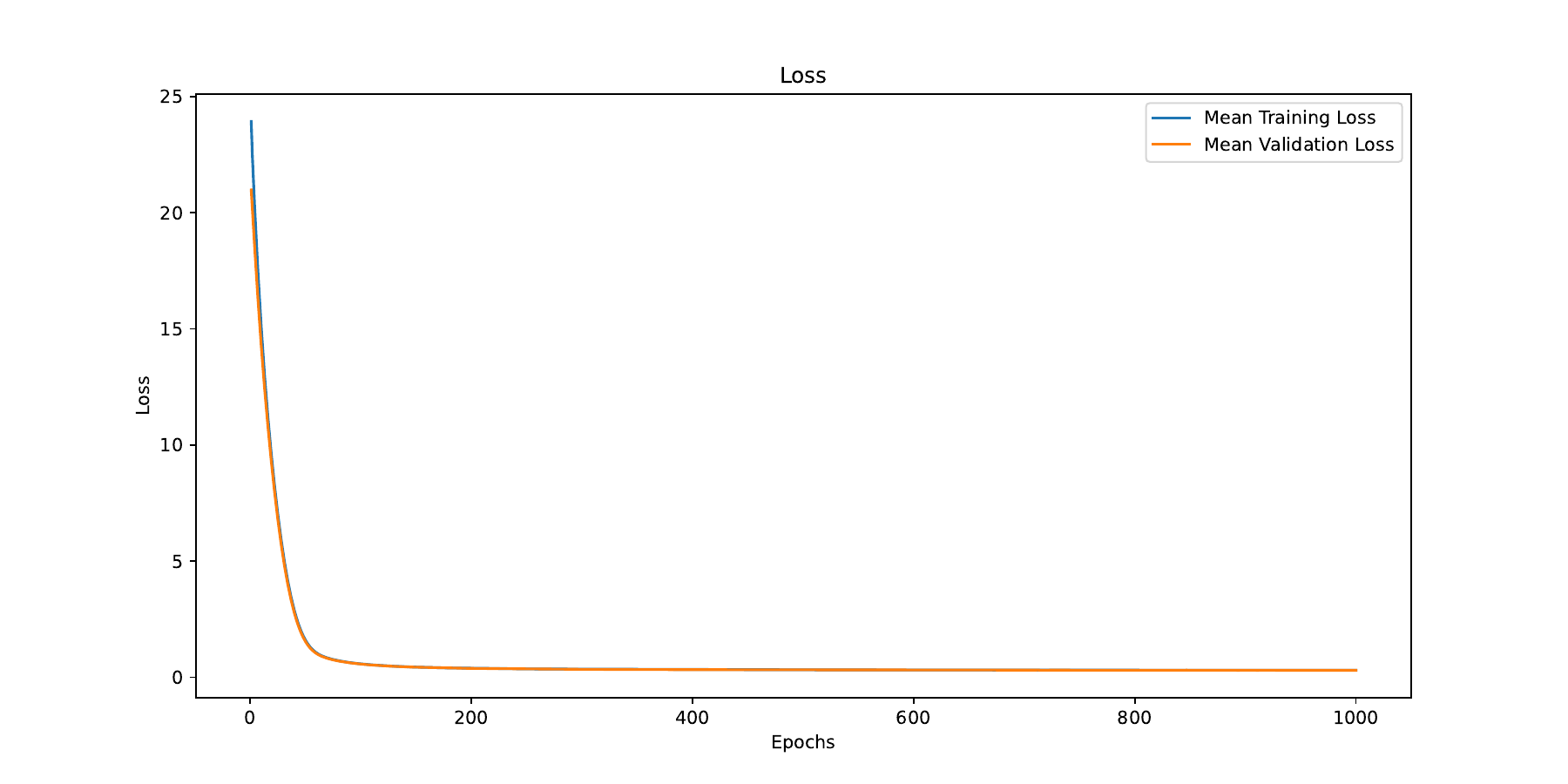}
        \caption{no UE, $N$-subjettiness}
        \label{fig:dnn_loss_nsub_noue}
    \end{subfigure}
    \hfill
    \begin{subfigure}[b]{0.49\textwidth}
        \centering
        \includegraphics[width=\textwidth]{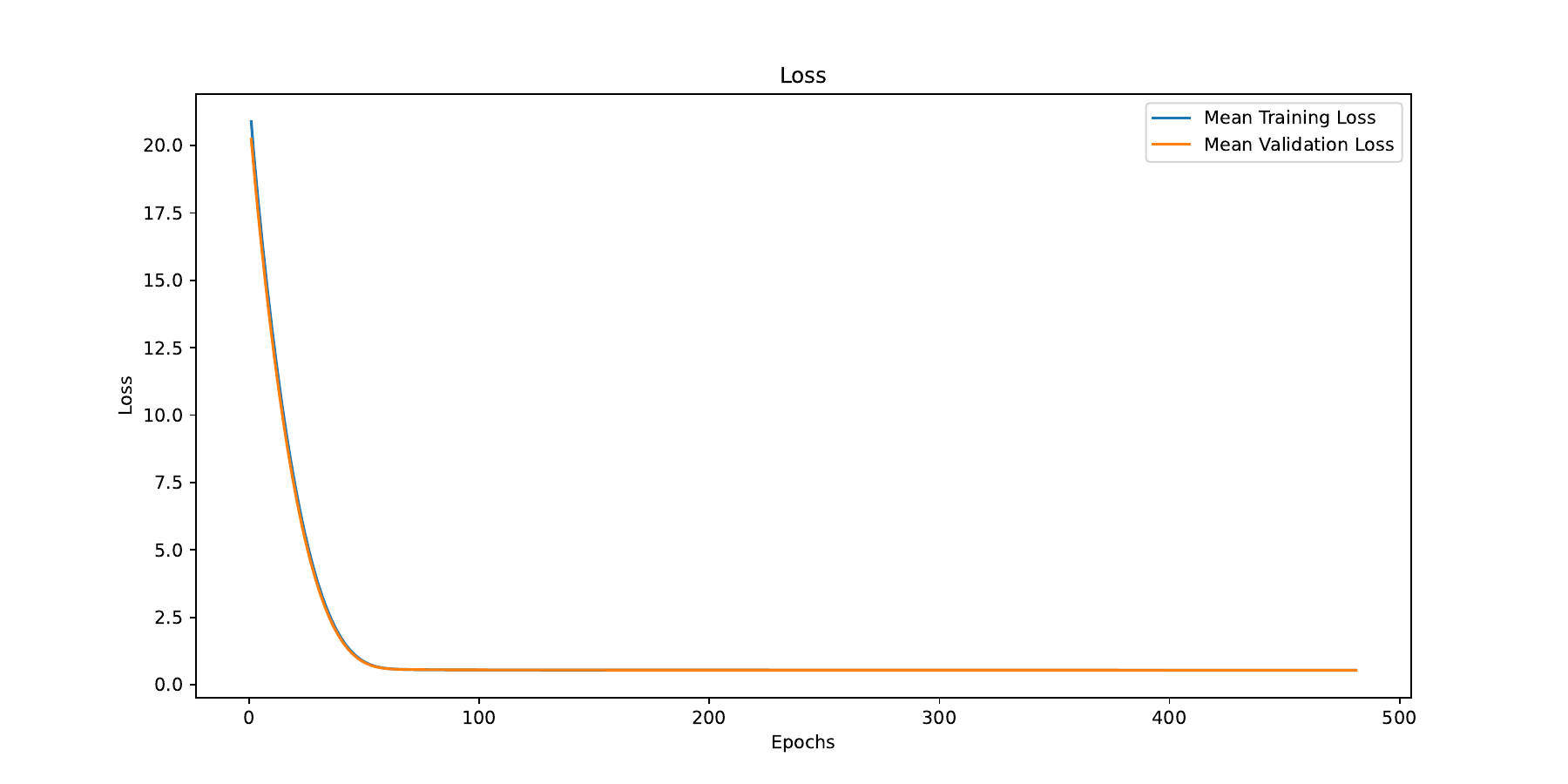}
        \caption{with UE, $N$-subjettiness}
        \label{fig:dnn_loss_nsub_ue}
    \end{subfigure}
    \begin{subfigure}[b]{0.49\textwidth}
        \centering
        \includegraphics[width=\textwidth]{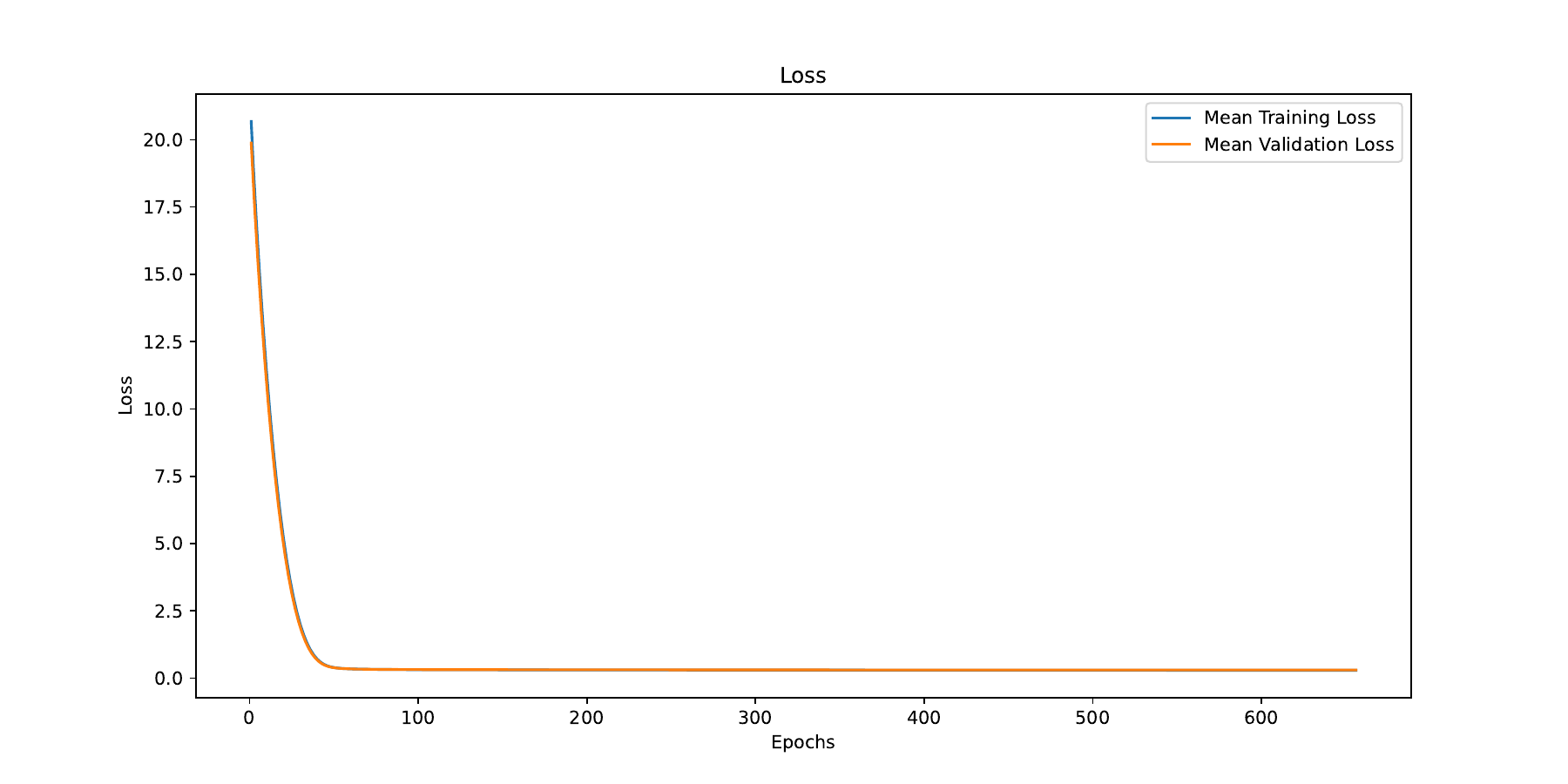}
        \caption{no UE, EFPs}
        \label{fig:dnn_loss_efp_noue}
    \end{subfigure}
    \hfill
    \begin{subfigure}[b]{0.49\textwidth}
        \centering
        \includegraphics[width=\textwidth]{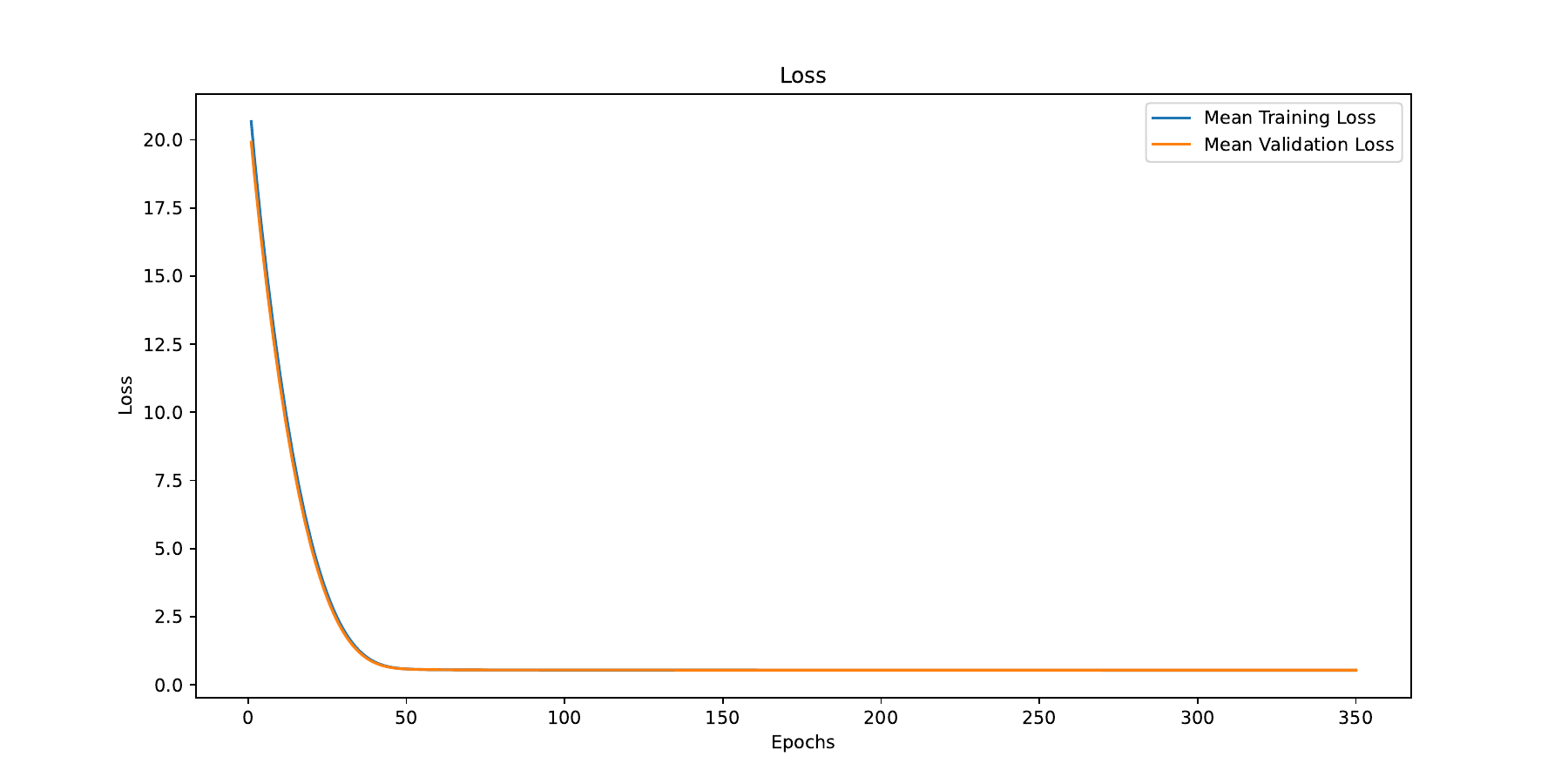}
        \caption{with UE, EFPs}
        \label{fig:dnn_loss_efp_ue}
    \end{subfigure}
    \begin{subfigure}[b]{0.49\textwidth}
        \centering
        \includegraphics[width=\textwidth]{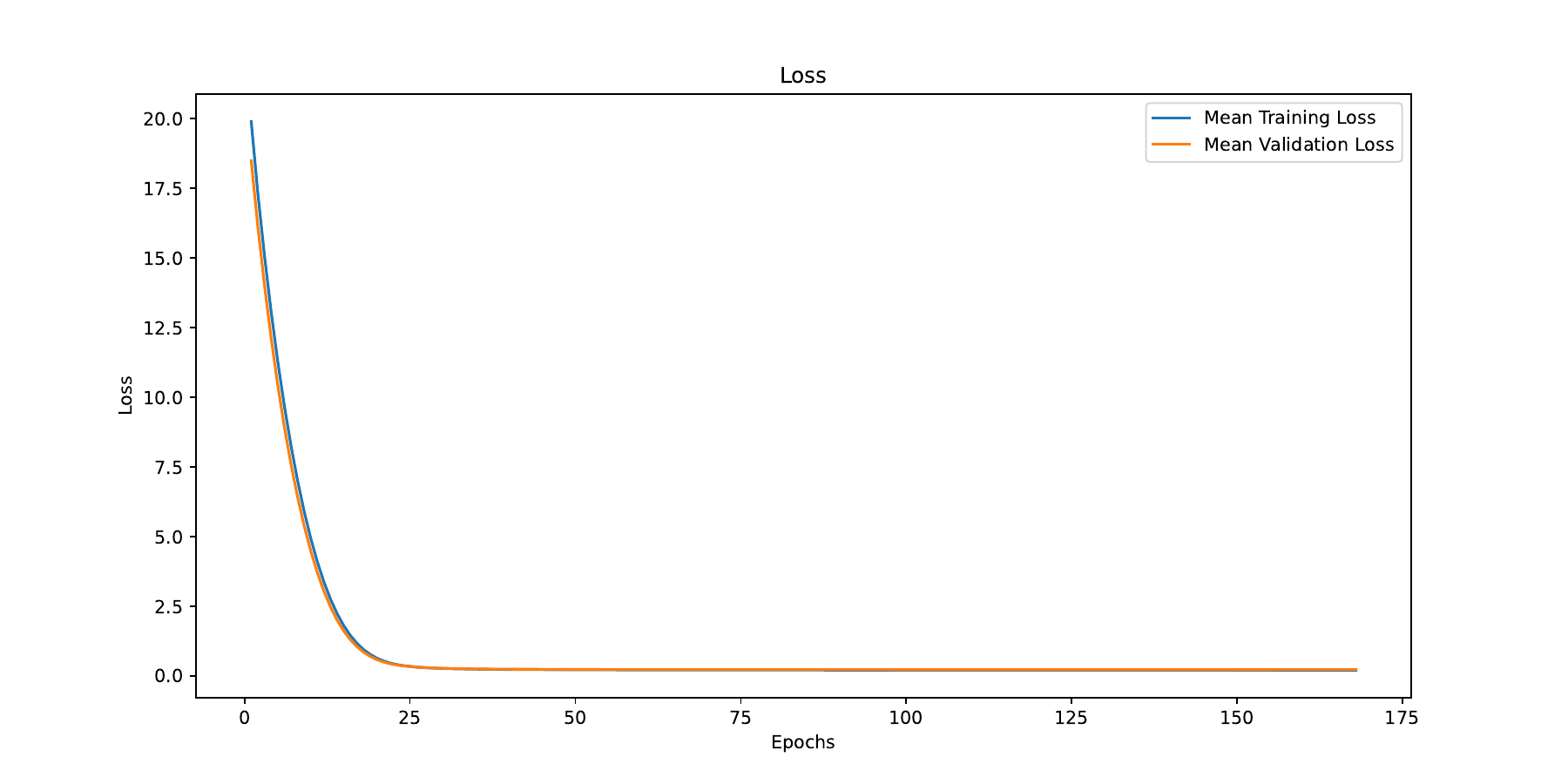}
        \caption{no UE, extended EFPs}
        \label{fig:dnn_loss_efpext_noue}
    \end{subfigure}
    \hfill
    \begin{subfigure}[b]{0.49\textwidth}
        \centering
        \includegraphics[width=\textwidth]{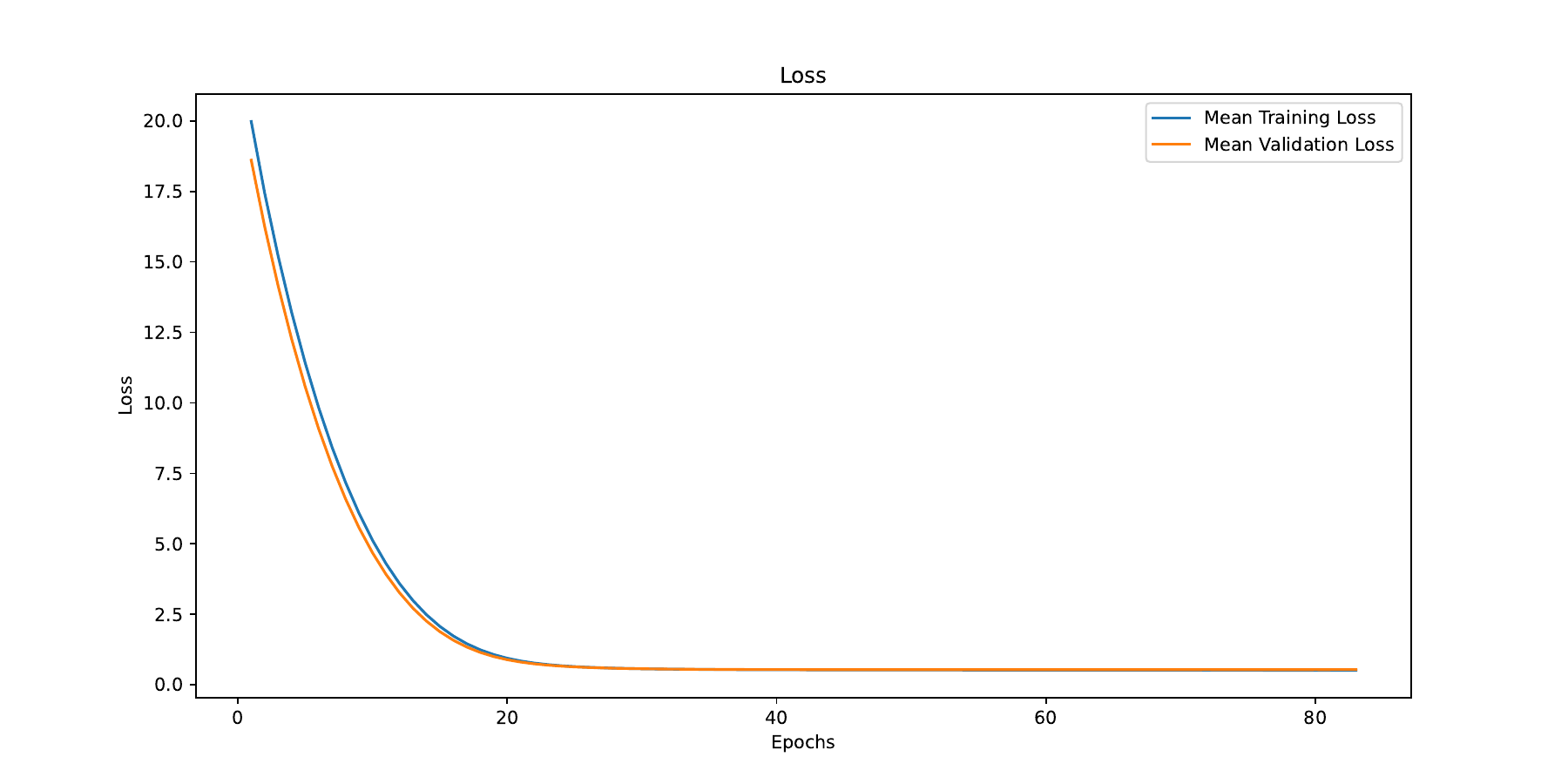}
        \caption{with UE, extended EFPs}
        \label{fig:dnn_loss_efpext_ue}
    \end{subfigure}
    \caption{Training/validation loss evolution for NNs across input sets and UE configurations.}
    \label{fig:dnn_losses_six}
\end{figure}

The training and validation losses for the non-linear baselines are shown in Fig.~\ref{fig:dnn_losses_six} and indicate stable convergence across all six configurations. For $N$-subjettiness (Figs.~\ref{fig:dnn_loss_nsub_noue},~\ref{fig:dnn_loss_nsub_ue}), EFPs (Figs.~\ref{fig:dnn_loss_efp_noue},~\ref{fig:dnn_loss_efp_ue}), and extended EFPs (Figs.~\ref{fig:dnn_loss_efpext_noue},~\ref{fig:dnn_loss_efpext_ue}), the curves decrease smoothly and plateau with no visible train-validation gaps, showing no signs of underfitting or overfitting. The very slightly higher validation floor with UE is consistent with the harder classification task under background contamination.

\begin{figure}[!htb]
    \centering
    \begin{subfigure}[b]{0.48\textwidth}
        \centering
        \includegraphics[width=\textwidth]{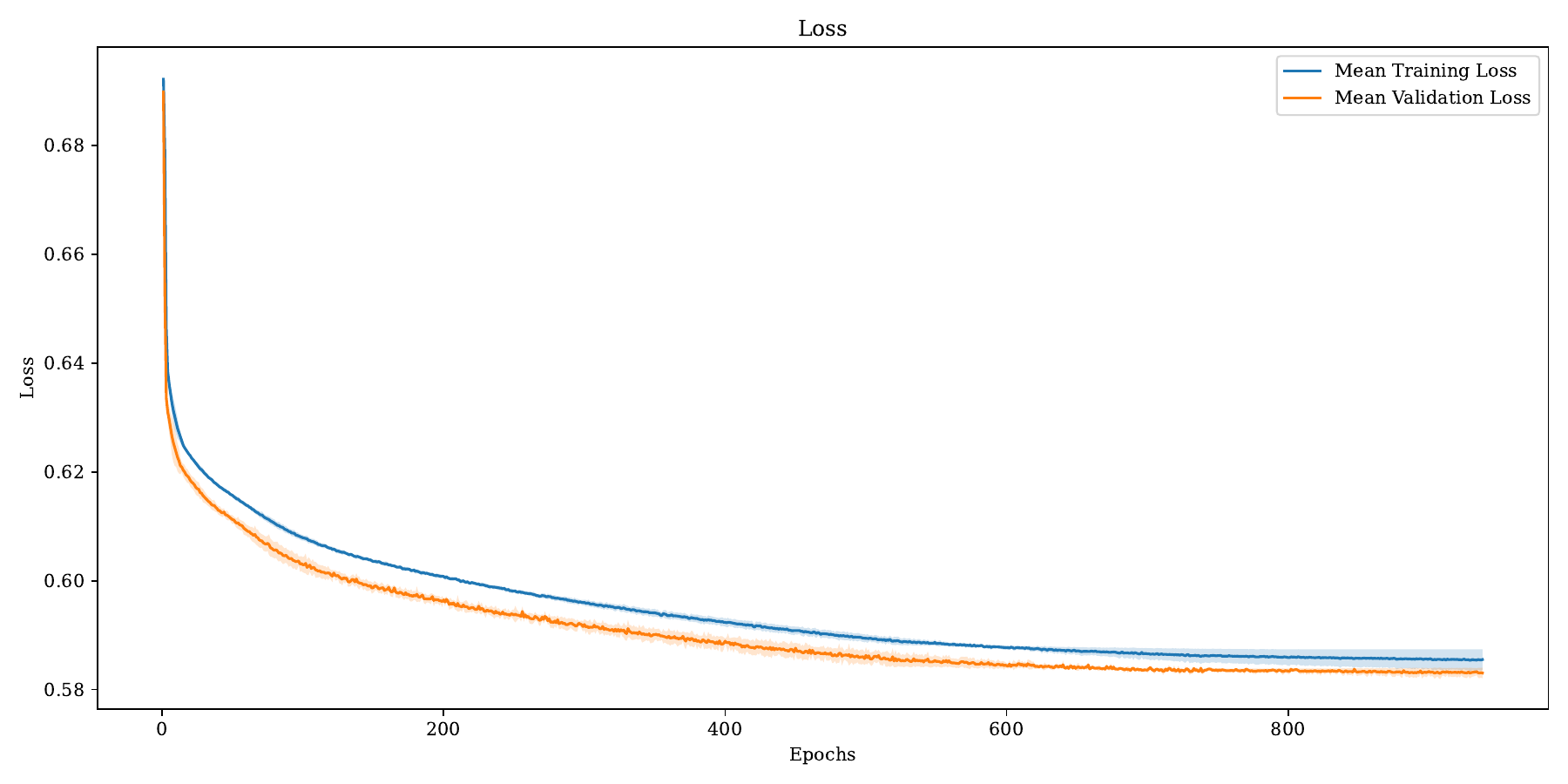}
        \caption{Vanilla}
        \label{fig:oefn_loss_vanilla}
    \end{subfigure}
    \hfill
    \begin{subfigure}[b]{0.48\textwidth}
        \centering
        \includegraphics[width=\textwidth]{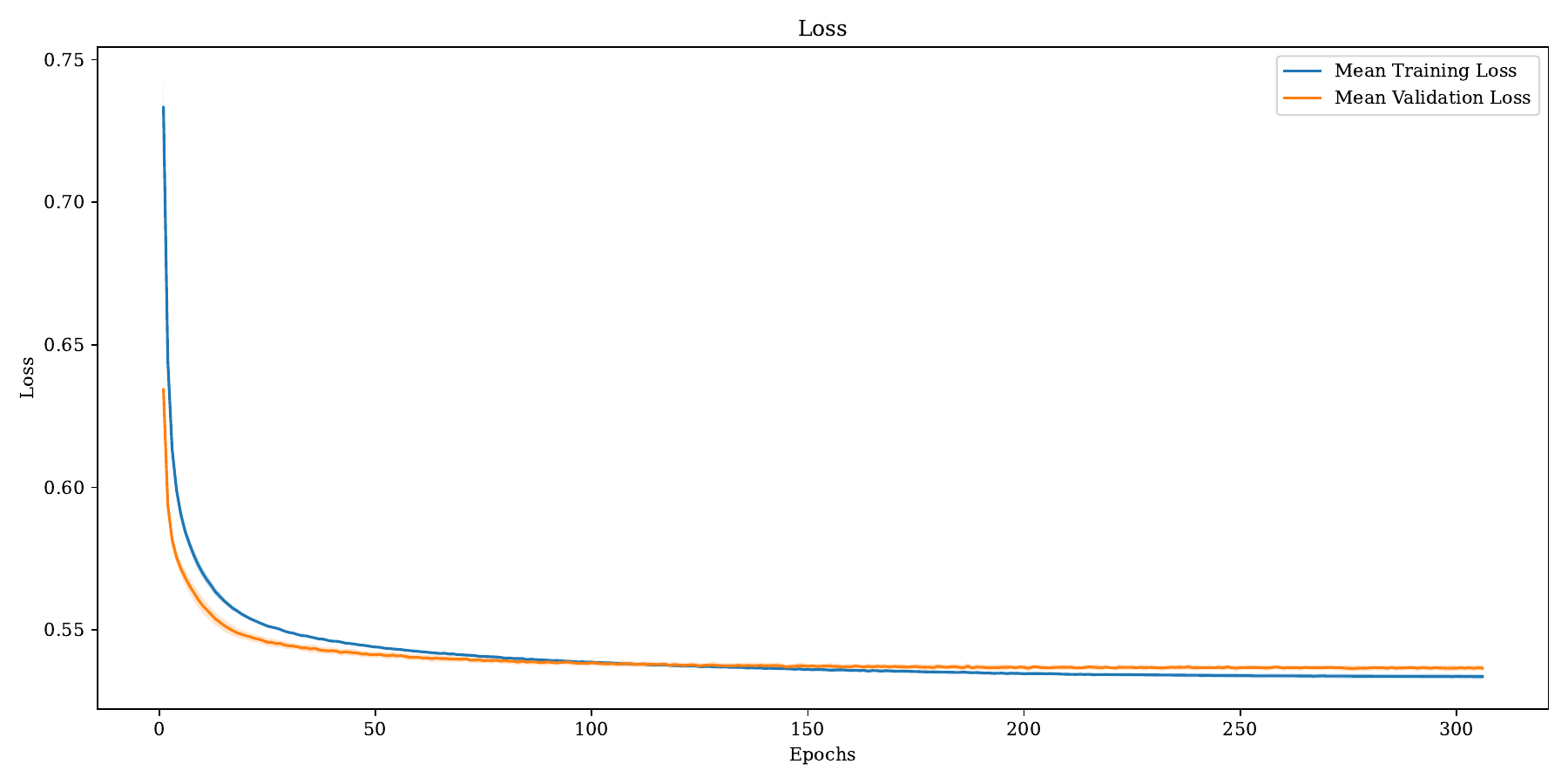}
        \caption{$N$-subjettiness}
        \label{fig:oefn_loss_nsub}
    \end{subfigure}
    \\
    \begin{subfigure}[b]{0.48\textwidth}
        \centering
        \includegraphics[width=\textwidth]{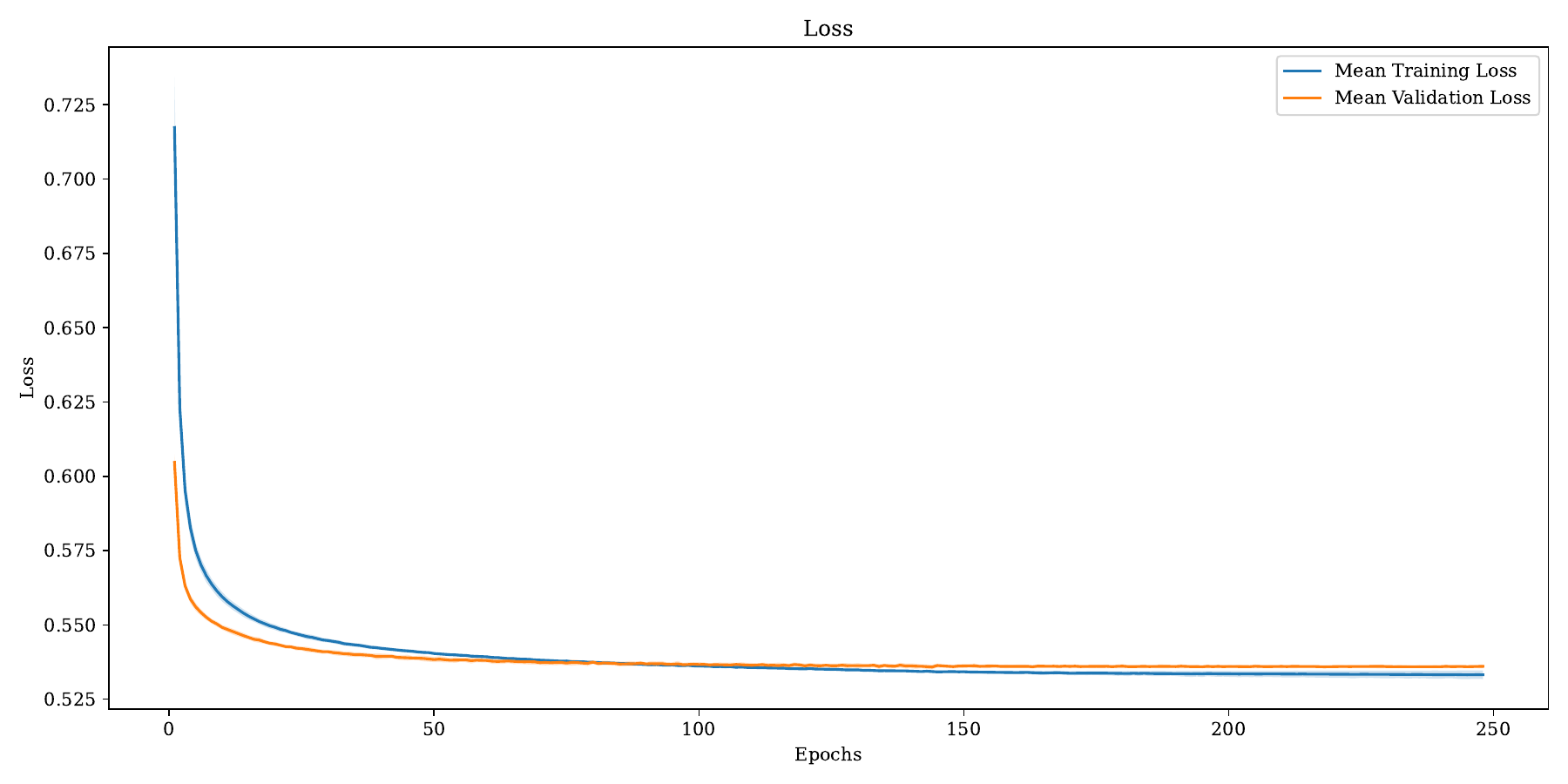}
        \caption{EFPs}
        \label{fig:oefn_loss_efp}
    \end{subfigure}
    \hfill
    \begin{subfigure}[b]{0.48\textwidth}
        \centering
        \includegraphics[width=\textwidth]{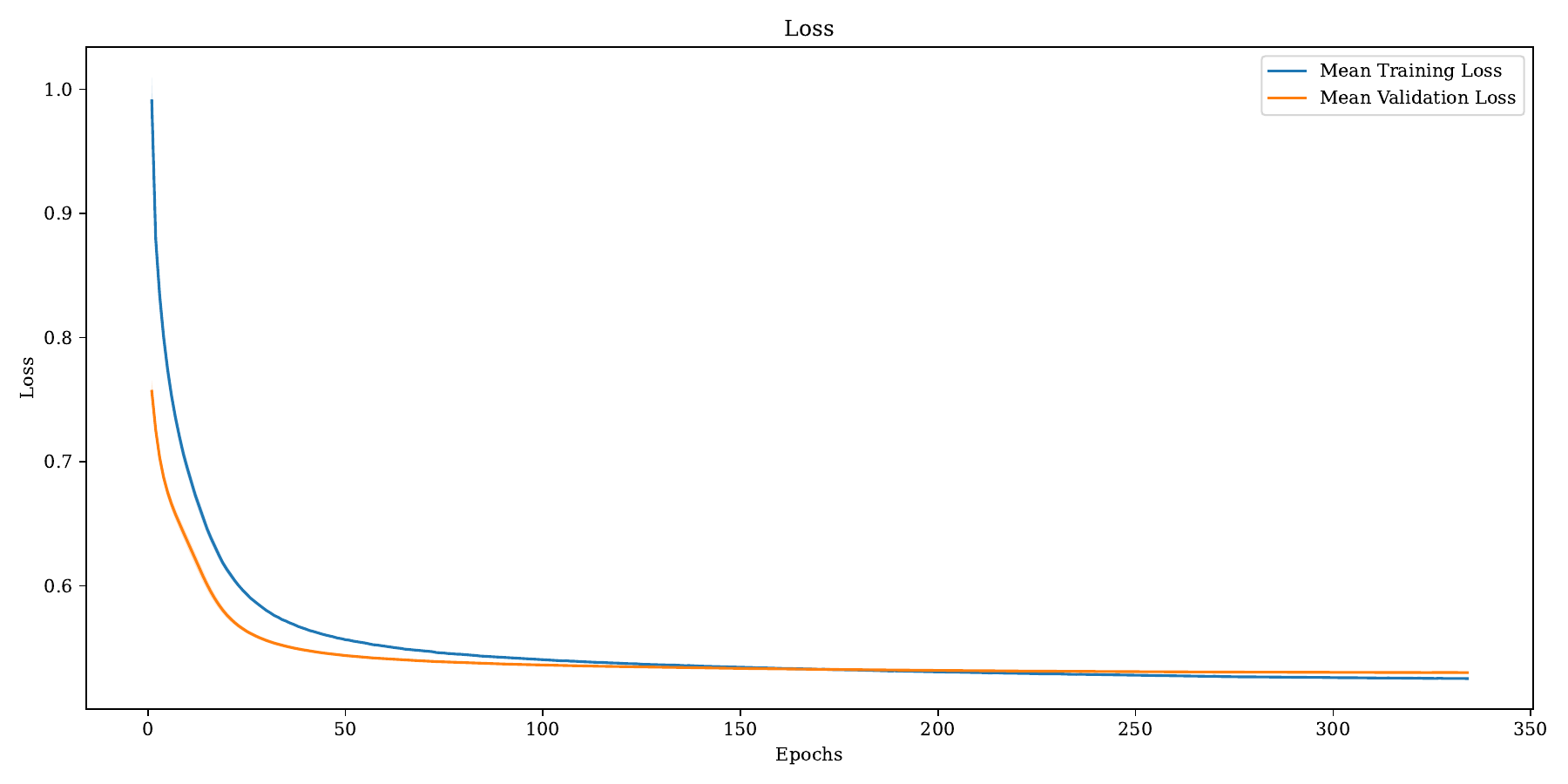}
        \caption{Extended EFPs}
        \label{fig:oefn_loss_efpext}
    \end{subfigure}
    \\
    \begin{subfigure}[b]{0.48\textwidth}
        \centering
        \includegraphics[width=\textwidth]{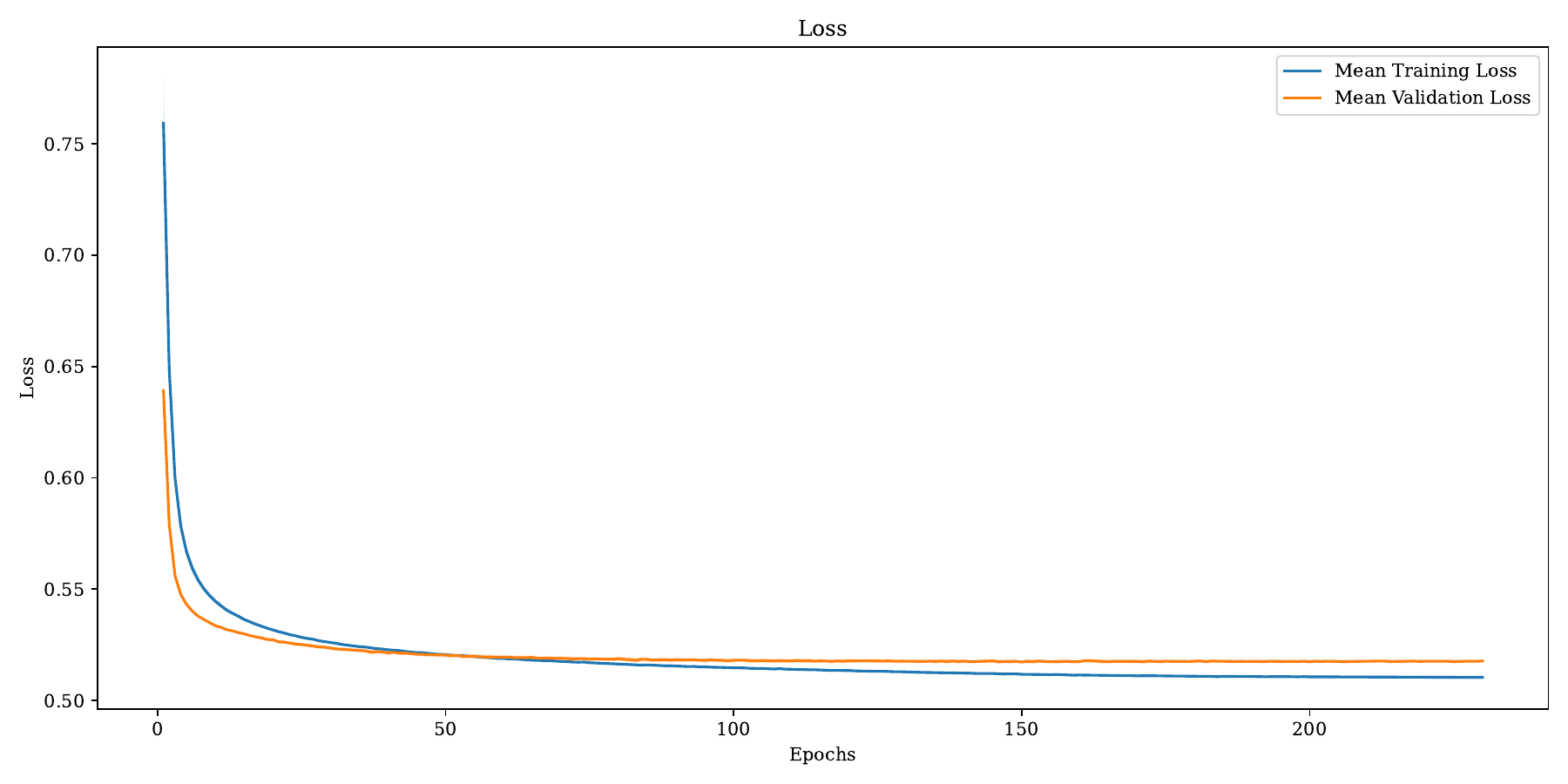}
        \caption{$N$-subjettiness + EFPs}
        \label{fig:oefn_loss_both}
    \end{subfigure}
    \hfill
    \begin{subfigure}[b]{0.48\textwidth}
        \centering
        \includegraphics[width=\textwidth]{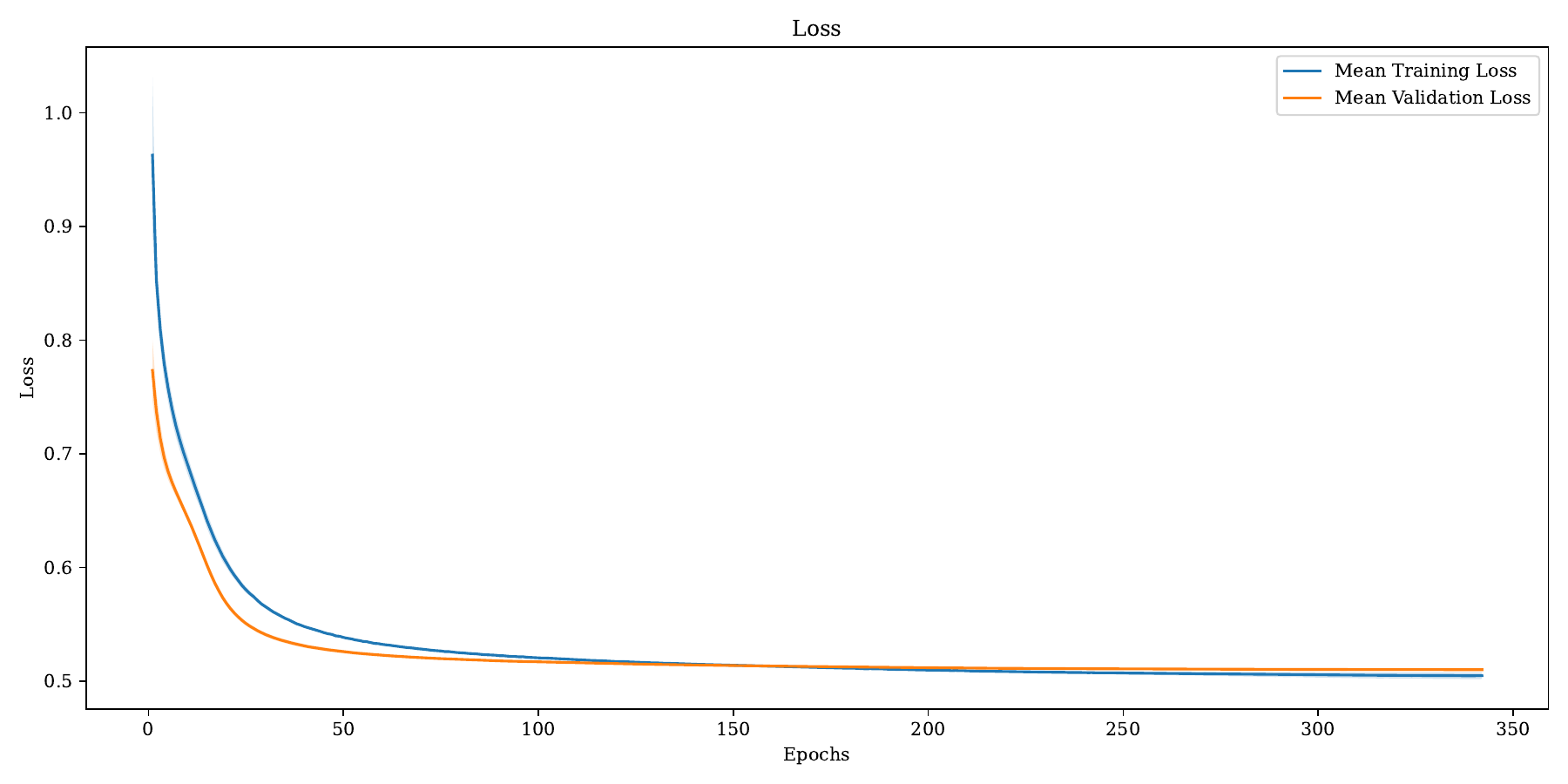}
        \caption{$N$-subjettiness + Extended EFPs}
        \label{fig:oefn_loss_both_ext}
    \end{subfigure}
    \caption{Training and validation losses for oEFNs under different augmentations.}
    \label{fig:oefn_losses}
\end{figure}

Training and validation losses for the oEFN models are shown in Fig.~\ref{fig:oefn_losses} and indicate stable optimization across all six configurations. The vanilla EFN curves (Fig.~\ref{fig:oefn_loss_vanilla}) and the augmented variants (Figs.~\ref{fig:oefn_loss_nsub}--\ref{fig:oefn_loss_both_ext}) decrease smoothly and plateau with negligible train-validation gaps. All losses show a non-existent or small gap between validation and training loss, showing no significant evidence of underfitting or overfitting.

\begin{figure}[!htbp]
    \centering
    \begin{subfigure}[b]{0.48\textwidth}
        \centering
        \includegraphics[width=\textwidth]{efn_best_model_mean_vs_std.png}
        \caption{Vanilla}
        \label{fig:oefn_cv_vanilla}
    \end{subfigure}
    \hfill
    \begin{subfigure}[b]{0.48\textwidth}
        \centering
        \includegraphics[width=\textwidth]{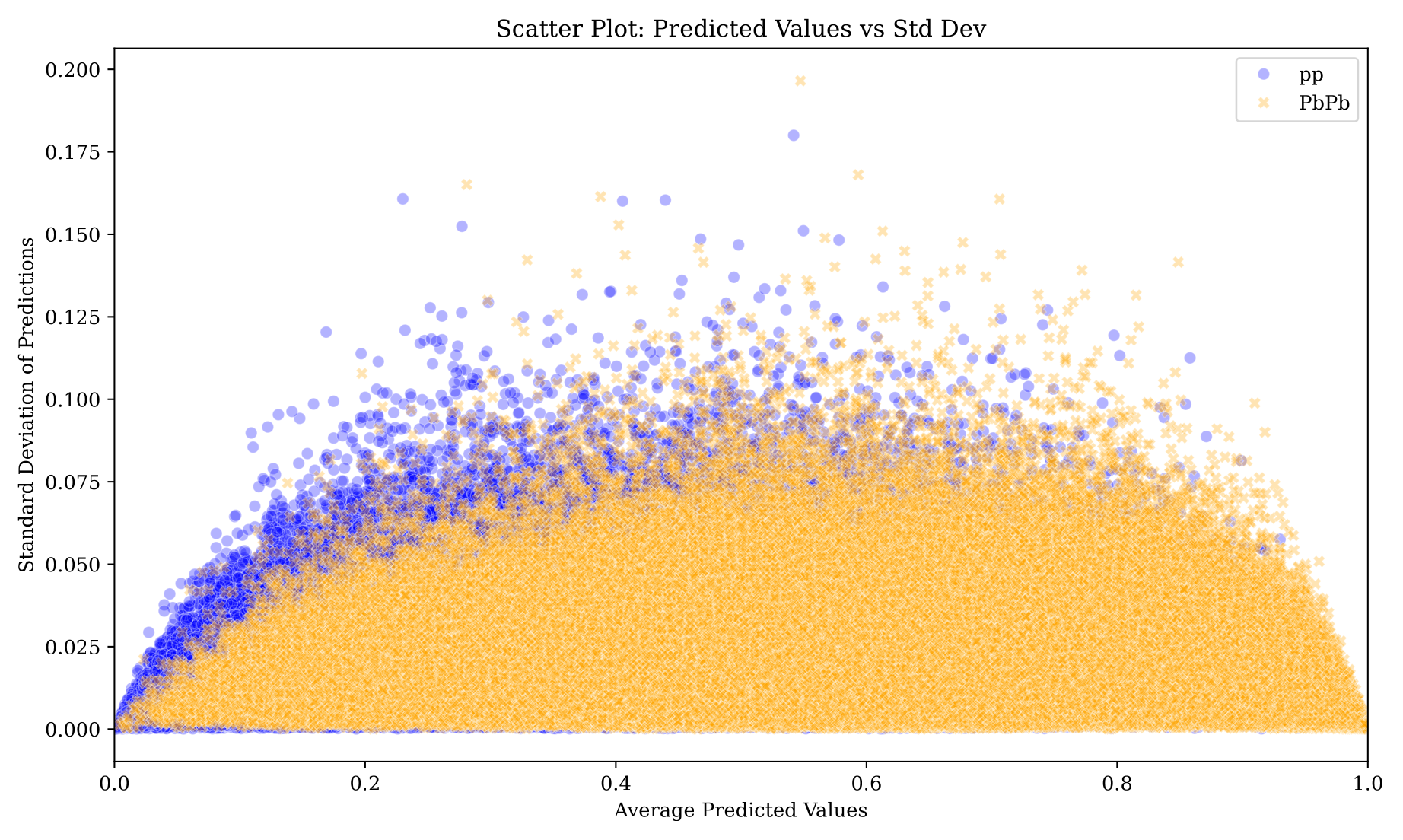}
        \caption{$N$-subjettiness}
        \label{fig:oefn_cv_nsub}
    \end{subfigure}
    \\
    \begin{subfigure}[b]{0.48\textwidth}
        \centering
        \includegraphics[width=\textwidth]{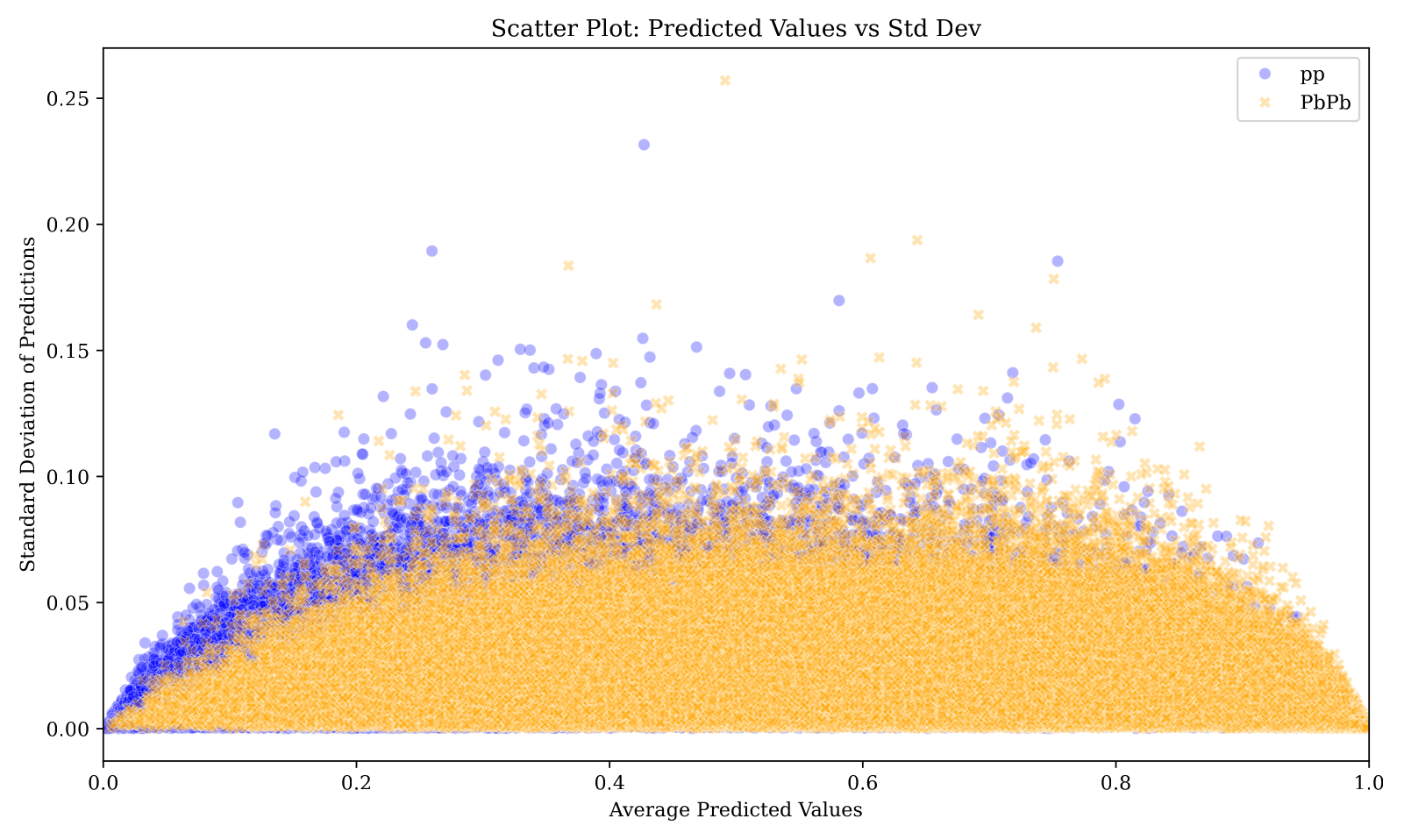}
        \caption{EFPs}
        \label{fig:oefn_cv_efp}
    \end{subfigure}
    \hfill
    \begin{subfigure}[b]{0.48\textwidth}
        \centering
        \includegraphics[width=\textwidth]{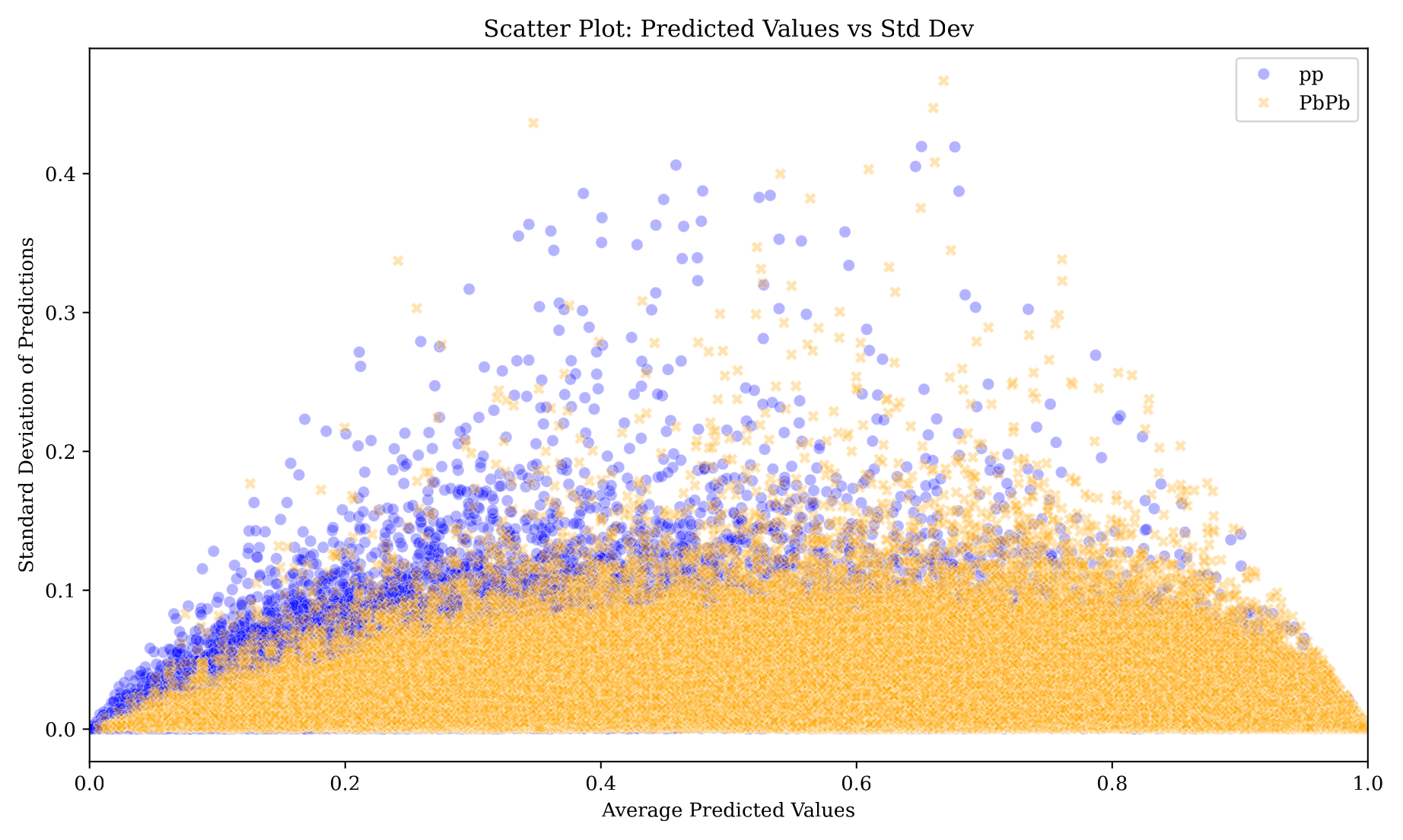}
        \caption{Extended EFPs}
        \label{fig:oefn_cv_efpext}
    \end{subfigure}
    \\
    \begin{subfigure}[b]{0.48\textwidth}
        \centering
        \includegraphics[width=\textwidth]{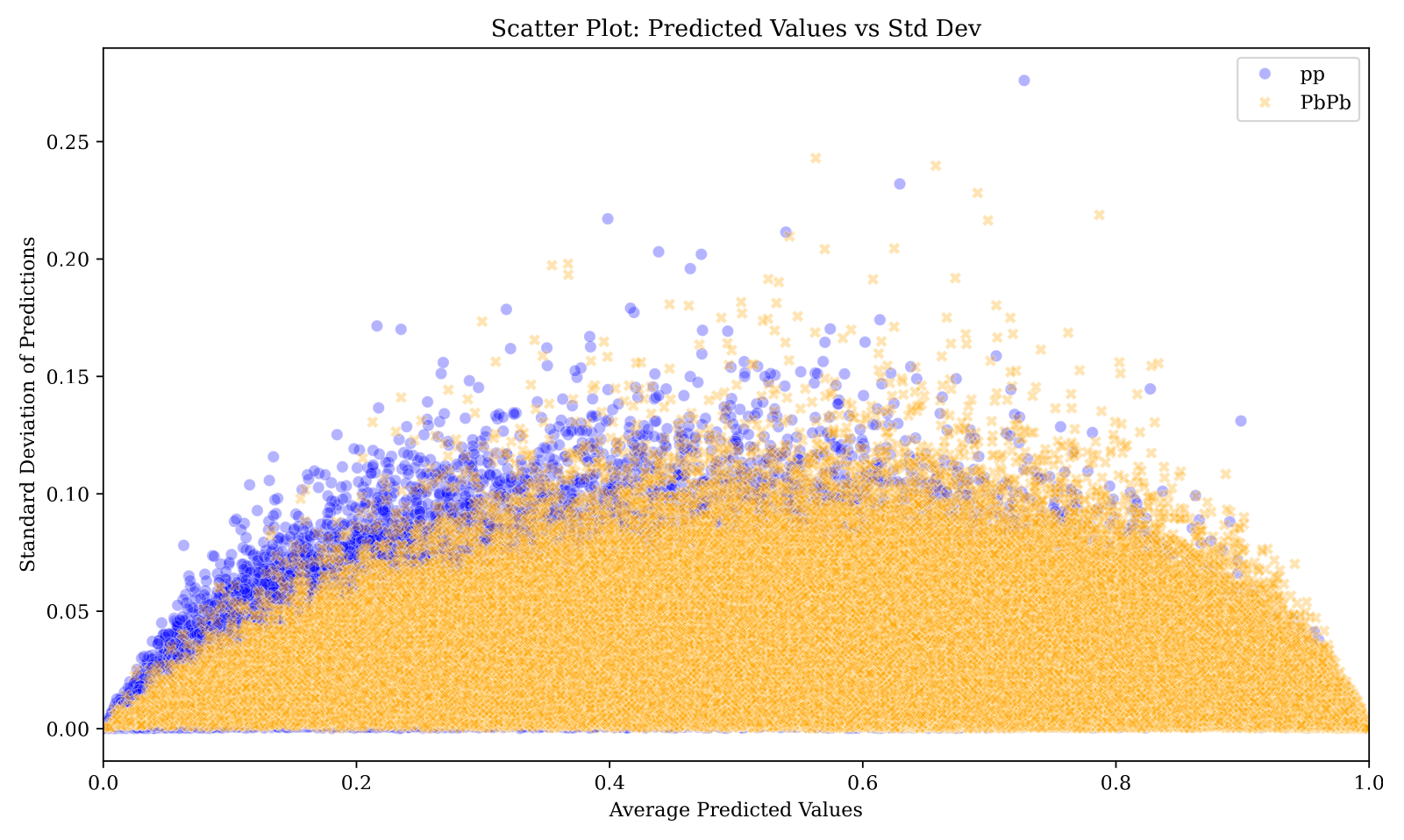}
        \caption{$N$-subjettiness + EFPs}
        \label{fig:oefn_cv_both}
    \end{subfigure}
    \hfill
    \begin{subfigure}[b]{0.48\textwidth}
        \centering
        \includegraphics[width=\textwidth]{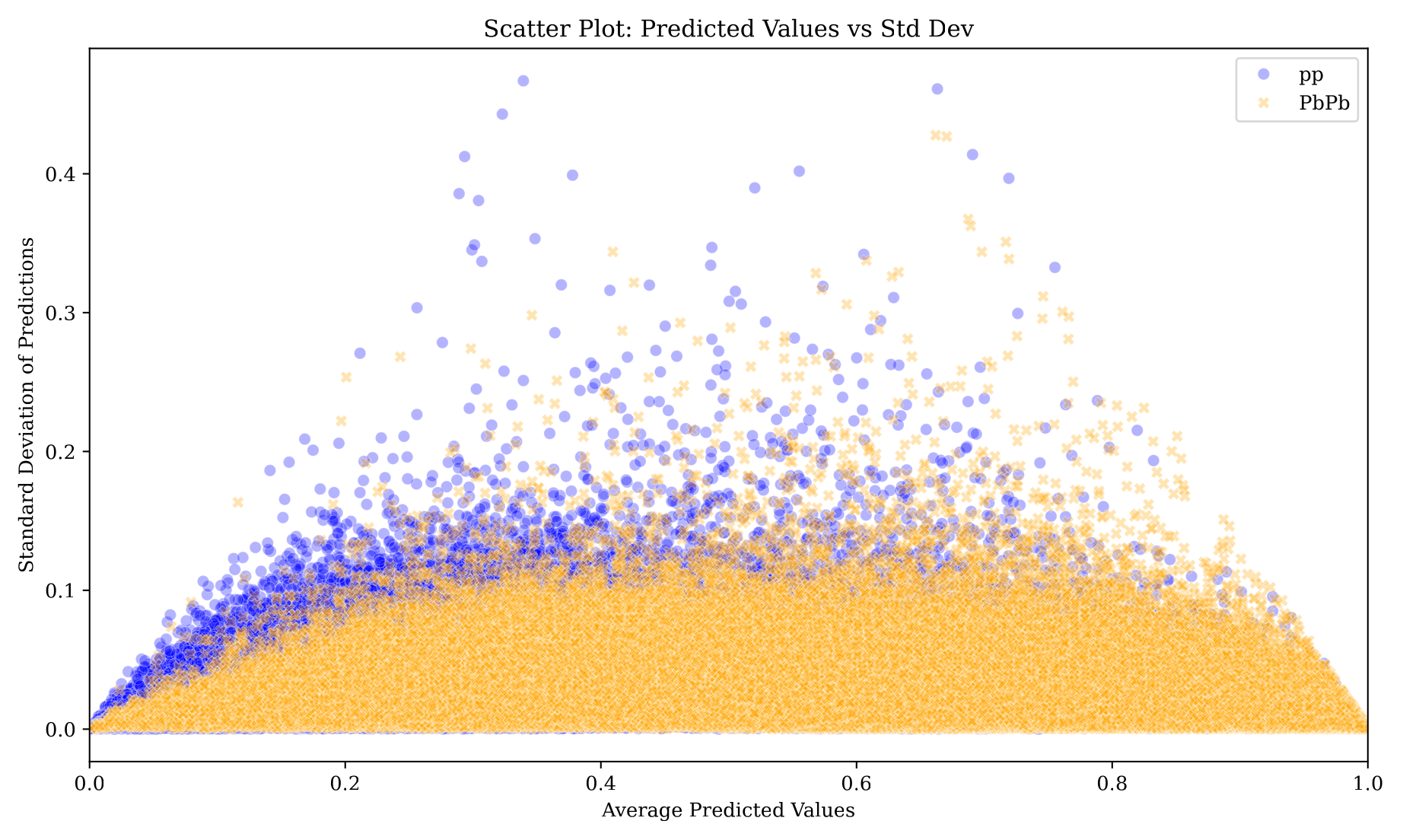}
        \caption{$N$-subjettiness + Extended EFPs}
        \label{fig:oefn_cv_both_ext}
    \end{subfigure}
    \caption{Cross-validation stability of oEFNs: mean vs.\ standard deviation across folds.}
    \label{fig:oefn_mean_std}
\end{figure}

Cross-validation stability, for the oEFN models, summarized by the taco plots in Fig.~\ref{fig:oefn_mean_std}, shows small fold-to-fold variability near classifier outputs close to 0 or 1 for all cases. Around intermediate scores, augmenting with $N$-subjettiness and/or standard EFPs reduces the standard deviation relative to the vanilla EFN (cf.~\ref{fig:oefn_cv_vanilla} vs.~\ref{fig:oefn_cv_nsub},~\ref{fig:oefn_cv_efp},~\ref{fig:oefn_cv_both}), indicating more stable decision boundaries. In contrast, using extended EFPs increases the variability in the central region for some data points (Figs.~\ref{fig:oefn_cv_efpext} and~\ref{fig:oefn_cv_both_ext}), in some ranges exceeding the vanilla level; this is consistent with expectation and suggests mild sensitivity to fold composition, even though overall AUC improves.

\phantomsection
\bibliographystyle{JHEP}
\bibliography{efjq}

\end{document}